\documentclass[
 reprint,
 showkeys,
superscriptaddress,
 amsmath,amssymb,
]{revtex4-1}

\usepackage{graphicx}
\usepackage{dcolumn}
\usepackage{bm}
\usepackage{siunitx}
\usepackage[utf8]{inputenc}
\usepackage{color,soul}
\usepackage{refstyle}

\begin{document}
\title{Tunable resistivity exponents in the metallic phase of epitaxial nickelates}
\author{Qikai Guo}
\email{q.guo@rug.nl}
\affiliation{Zernike Institute for Advanced Materials, University of Groningen, The Netherlands}
\author{Saeedeh Farokhipoor}
\affiliation{Zernike Institute for Advanced Materials, University of Groningen, The Netherlands}
\author{C{\'e}sar Mag{\'e}n} 
\affiliation{Instituto de Ciencia de Materiales de Arag{\'o}n (ICMA), CSIC-Universidad de Zaragoza, 50009 Zaragoza, Spain}
\author{Francisco Rivadulla}
\affiliation{Departamento de Qu{\'i}mica-F{\'i}sica, Universidade de Santiago de Compostela, Santiago de Compostela 15782, Spain}
\author{Beatriz Noheda}
\email{b.noheda@rug.nl}
\affiliation{Zernike Institute for Advanced Materials, University of Groningen, The Netherlands}
\affiliation{CogniGron center, University of Groningen, The Netherlands}

\date{\today}

\begin{abstract}
We report a detailed analysis of the electrical resistivity exponent of thin films of NdNiO\textsubscript{3} as a function of epitaxial strain. Strain-free thin-films show a linear dependence of the resistivity \textit{versus} temperature, consistent with a classical Fermi gas ruled by electron-phonon interactions. In addition, the apparent  temperature exponent, \textit{n}, can be tuned with the epitaxial strain between \textit{n}= 1 and \textit{n}= 3. We discuss the critical role played by quenched random disorder in the value of \textit{n}. Our work shows that the assignment of Fermi/Non-Fermi liquid behaviour based on experimentally obtained resistivity exponents requires an in-depth analysis of the degree of disorder in the material. \\ 

\end{abstract}
 
\maketitle

The tunable resistivity of materials undergoing a metal-insulator transitions (MIT) holds great promise for resistive switching applications, such as adaptable electronics and cognitive computing \cite{scherwitzl2010electric, driscoll2009phase, mcleod2017nanotextured, ha2011metal, ha2014neuromimetic, shi2013correlated, wang2017electrochemically}. However, a complete understanding of the metallic phase in these strongly correlated electron systems is still one of the central open problems in condensed matter physics \cite{emery1995superconductivity,zaanen2019}. 

Electronic transport is generally explained by means of Boltzmann's theory, which considers a fluid of free quasi-particles that scatter occasionally. In normal metals, the resistivity increases linearly with temperature as electrons are more strongly scattered by lattice vibrations. At low temperatures, weak interactions between electrons can significantly affect the electrical properties and give rise to a \textit{T}\textsuperscript{2} dependence of resistivity, according to Landau's Fermi liquid (FL) theory \cite{landau1956multiplicative}. Therefore, the scaling exponent  of the power law term of the resistivity as a function of temperature (\textit{n}) is often used to infer the type of interactions ruling the metal state. In materials with strong electron-electron interactions and undergoing ordering phenomena, other exponents (\textit{n}$\neq $ 1, 2) are usually observed, being the physics behind this so-called “Non-Fermi liquid” (NFL) behaviour \cite{stewart2001non, schofield1999non, rivadulla2007vo} a subject of active discussion \cite{stemmer2018non, keller2015universal, kasahara2010evolution, lee2018recent}. 

 Among strongly correlated electron materials, nickelates (RENiO$_3$, with RE denoting a trivalent rare earth element) present a very interesting case. They have attracted attention due to their MIT \cite{imada1998metal} and the possibility to tune it using different RE elements or by epitaxial strain \cite{medarde1997structural,catalan2000transport,catalan2000metal,catalan2008progress,middey2016physics,catalano2018rare}. Bad metallic behaviour in nickelates has also been claimed\cite{jaramillo2014origins}. Different models for the origin of the MIT have been put forward, based on either positive or negative charge transfer as responsible for the insulating state \cite{zaanen1985band,mizokawa1991origin,alonso2000room,khomskii2001unusual,mazin2007charge, medarde2009charge,park2012site,johnston2014charge,varignon2017complete}. The negative charge transfer model supports the bond disproportionation picture and is strongly supported by recent experiments \cite{bisogni2016ground,green2016bond,shamblin2018experimental}. Independent from the exact microscopic picture, the origin of the MIT is a cooperative lattice distortion that reduces the symmetry from a high-temperature orthorhombic phase to a low-temperature monoclinic phase, involving two Ni sites, with the associated need for cooperative accommodation of different Ni-O bond-lengths \cite{mercy2017structurally}. Remarkably, it has been reported that eliminating the MIT in nickelates by orbital engineering would give rise to a superconducting state \cite{hansmann2009turning}, with a very recent experimental achievement in this direction \cite{danfeng2019superconductivity}. It becomes, then, important to have an accurate picture of the relevant electron interactions in the intermediate and low temperature regimes, just before the MIT takes place. However, despite the vast amount of recent works, the metallic behaviour of the nickelates is not yet fully understood.
 
\begin{figure*}
\includegraphics[width=0.7\textwidth]{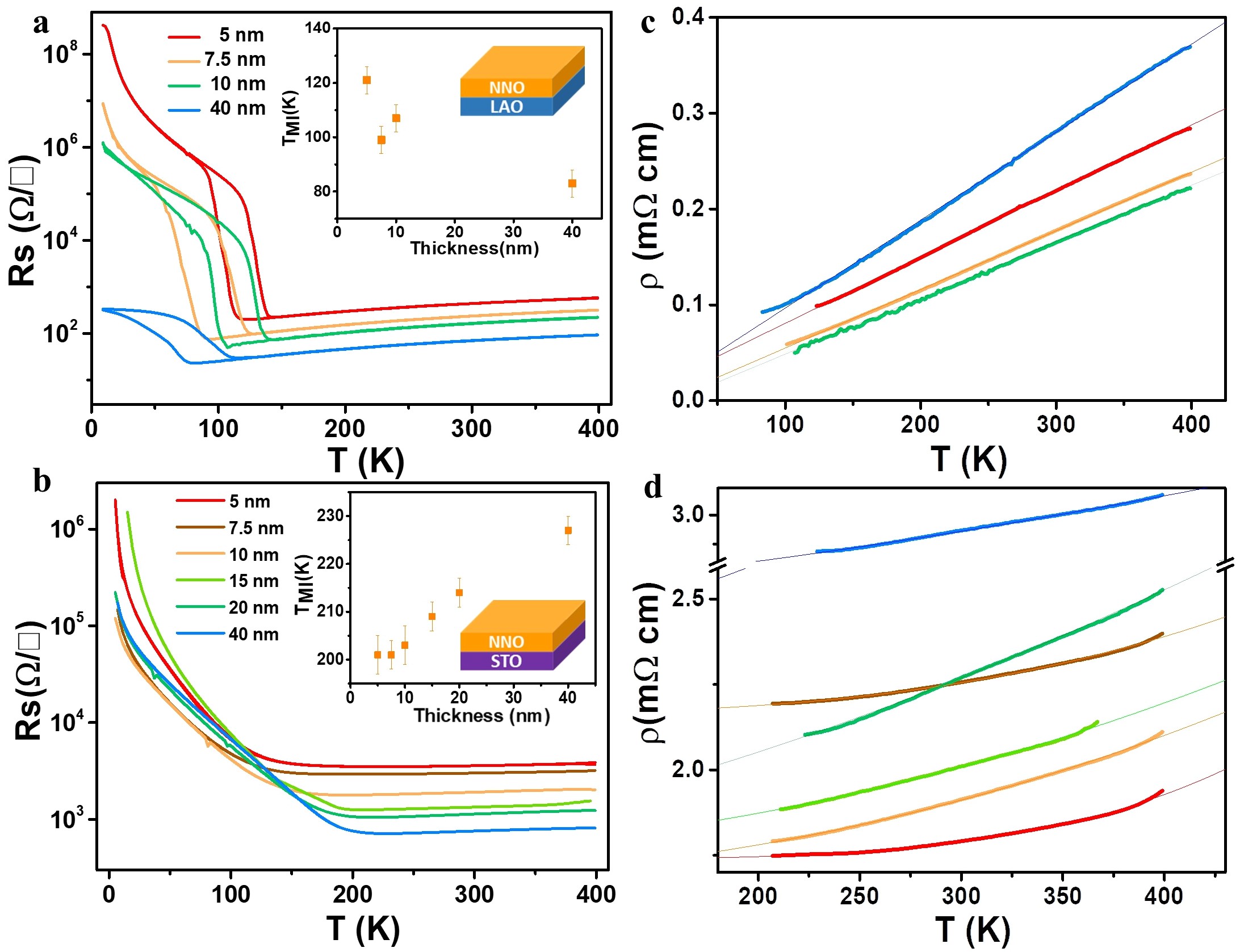}
\caption{Temperature dependence of the sheet resistance both during cooling and heating for NNO thin films grown on (a) LAO and (b) STO substrates with different thickness. Thermal hysteresis is only present in the films on LAO. The resistivity as a function of temperature in the metallic phase of NNO thin films grown on (c)  LAO and (d) STO substrates with different thickness. The thin solid lines are fits using Eq. (1).}
\label{fig:resistivity}
\end{figure*}

 In nickelates, different \textit{n}-exponents of the resistivity as a function of temperature have been reported \cite{blasco1994structural, liu2013heterointerface, jaramillo2014origins, mikheev2015tuning, kobayashi2015pressure, yadav2018influence, phanindra2018terahertz,stemmer2018non}. Linear dependence with temperature has been measured in the whole Nd$_{x}$La$_{1-x}$NiO$_{3}$ series in ceramic pellets\cite{blasco1994structural}. Liu et al. \cite{liu2013heterointerface} obtained \textit{n}= 5/3 and \textit{n}= 4/3 for NdNiO\textsubscript{3} (NNO) films under compressive strain, while Mikheev et al. reported a crossover between FL (\textit{n}=2) and NFL (\textit{n}=5/3) in NNO films with varying epitaxial strain \cite{mikheev2015tuning}. The need of an empirical parallel resistor model to introduce the effect of the saturation resistivity rises questions about the interpretation of the \textit{apparent} (experimentally obtained) exponents, as discussed by Hussey \textit{et al.} \cite{hussey2004universality}.

\begin{figure}
\includegraphics[width=.4\textwidth]{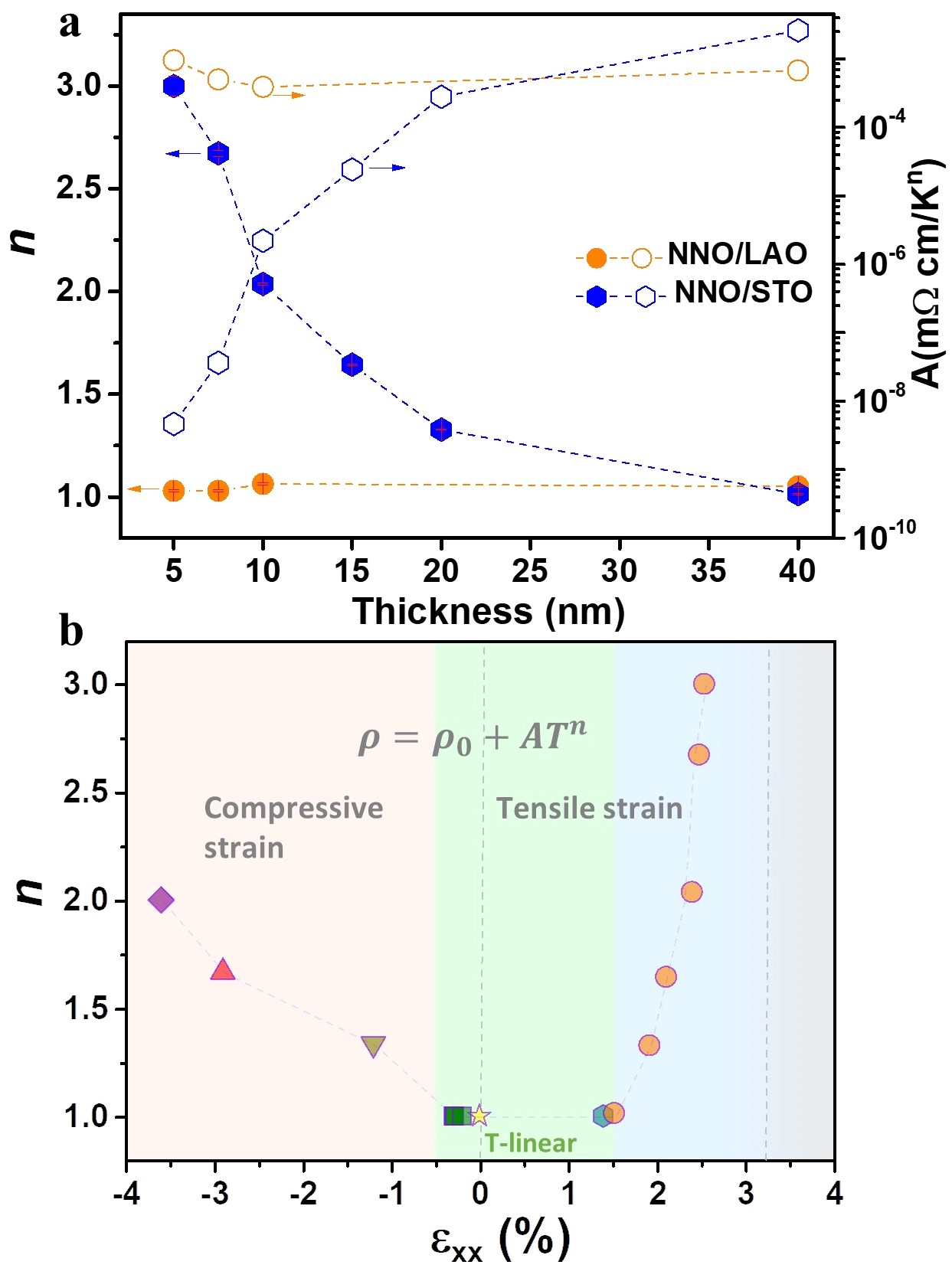}
\caption{(a) Power law exponents (\textit{n}) and A-coefficients from eq. (1), extracted from the fits in (c) and (d) as a function of film thickness. (b) Scaling exponent \textit{n} as a function of in-plane strain $\varepsilon$\textsubscript{xx}. The data are for films grown on different substrates: LAO (squares), NGO (hexagon) and STO (circles). The highly tensile region shadowed in grey denotes the insulating state observed for the films on DSO. In addition, we also plot \textit{n} of bulk NNO \cite{blasco1994structural} (star), as well as that for epitaxial NNO films under compressive strain reported by Liu et al. \cite{liu2013heterointerface} (triangles) and Mikheev et al. \cite{mikheev2015tuning} (rhombus).}
\label{fig:fitparameters}
\end{figure}

Here, we report the evolution of the resistivity exponent of NdNiO\textsubscript{3} under different degrees of epitaxial strain. Strain-free (bulk like) thin films show a linear temperature dependence of the resistivity (\textit{n}=1). The combined effect of epitaxial strain and random disorder produces a continuous departure from \textit{n}=1, in agreement with recent theoretical work by Patel \textit{et al}. \cite{patel2017non}.

Crystalline NNO films have been grown by pulsed laser deposition (PLD) on LaAlO$_3$ (LAO), NdGaO$_3$ (NGO), SrTiO$_3$ (STO) and DyScO$_3$ (DSO) substrates, using a single-phase ceramic target \cite{preziosi2017reproducibility}, as described in the supplemental Material (SM). Perovskite NNO possesses an orthorhombic structure with a pseudo-cubic lattice parameter of 3.807 $\si{\angstrom}$, which is slightly larger than that of the LAO substrate (3.790 $\si{\angstrom}$). Thus, the films on LAO are expected to be subjected to small compressive strain. On the contrary, the films grown on NGO (3.858 $\si{\angstrom}$), STO (3.905 $\si{\angstrom}$) and DSO (3.955 $\si{\angstrom}$) substrates should experience increasing tensile strain. Figure S1 (see SM) shows the typical atomic force microscope (AFM) topography image of a 5 nm NNO film grown on a LAO substrate (NNO/LAO), showing that the atom-high steps from the substrate are still visible after the deposition of the film. \textit{In-situ} high energy electron diffraction (RHEED) intensity oscillations recorded during the film growth indicate that at least the first 13 layers ($\sim$ 5 nm) of NNO film are deposited atomic-layer by atomic-layer (see Fig. S1(a) in SM for NNO/LAO and NNO/STO films). The crystalline quality and strain state of the NNO films with different thickness and on different substrates was determined by X-ray diffraction (for details see SM and later discussions).

Figure \ref{fig:resistivity}(a) and (b) shows the sheet resistance of NNO films grown on LAO and STO substrates, respectively, as a function of temperature (for the measurement details see SM). The NNO films grown on LAO substrates (under small compressive strain) exhibit a sharp MIT and a pronounced thermal hysteresis. On the contrary, the hysteresis is fully suppressed in the NNO/STO films, in agreement with previous reports \cite{scherwitzl2010electric}. The replacement of the first order transition by a continuous, percolative-like metal-insulator transition is consistent with the presence of quenched random disorder in the films grown on STO \cite{Salamon2002Colossal}. This interpretation is supported by a higher resistivity and a smaller residual resistivity ratio in these films compared to those grown on LAO. A further distinction is observed in the evolution of the metal-insulator transition temperature (\textit{T}\textsubscript{MI}) as a function of thickness. The trend observed in thin films grown on both substrates (see insets to Fig. 1(a) and (b)) has been attributed to the opposite alteration of out-of-plane Ni-O-Ni angle in response to different sign of the epitaxial strain \cite{catalano2018rare}. 

Like in most of metals, the electrical resistivity in the metallic state of nickelates can be fitted using a power law:

\begin{equation}\label{3}
 \rho(T)=\rho(0)+AT^n
\end{equation}

 where \textit{A} is a coefficient related to the strength of electron scattering and \textit{n} is the apparent power law exponent. As shown in Fig. \ref{fig:resistivity}(c), the metallic resistivity of all NNO films grown on LAO substrates in the measured temperature range (from $T$\textsubscript{MI}$\sim$ 100 K to 400 K) can be well described with a linear temperature dependence (\textit{n}=1.00$\pm$ 0.01), independent of film thickness. This temperature dependence has been observed in other systems, ranging from cuprates to heavy fermions, in spite of their different mechanisms of electron scattering \cite{bruin2013similarity}. What they have in common, however, is a constant scattering rate per kelvin ($\approx k_B/\hbar$), indicating that the excitations responsible for scattering are governed only by temperature. On the other hand, in the case of NNO/STO films (Fig. \ref{fig:resistivity}(d)), the  temperature-resistivity scaling of films with different thickness deviates from linearity, showing the departure from this intrinsic mechanism. The values of \textit{n} and A-coefficients in both NNO/LAO and NNO/STO systems are shown, as a function of thickness, in Fig. \ref{fig:fitparameters}(a) (for details on the determination of \textit{n}, see Fig. S2 in SM). Interestingly, \textit{n} shows a clear evolution with thickness in the NNO/STO films: \textit{n} decreases with increasing NNO/STO film thickness from a value of \textit{n}= 3.00$\pm$ 0.05 for a 5 nm film to an apparent linear dependence (\textit{n}= 1.01$\pm$ 0.01) for the thickest film (40 nm). To understand this behaviour we turn to an in-depth structural characterization of the films.

\begin{figure}
\centering
\includegraphics[width=0.5\textwidth]{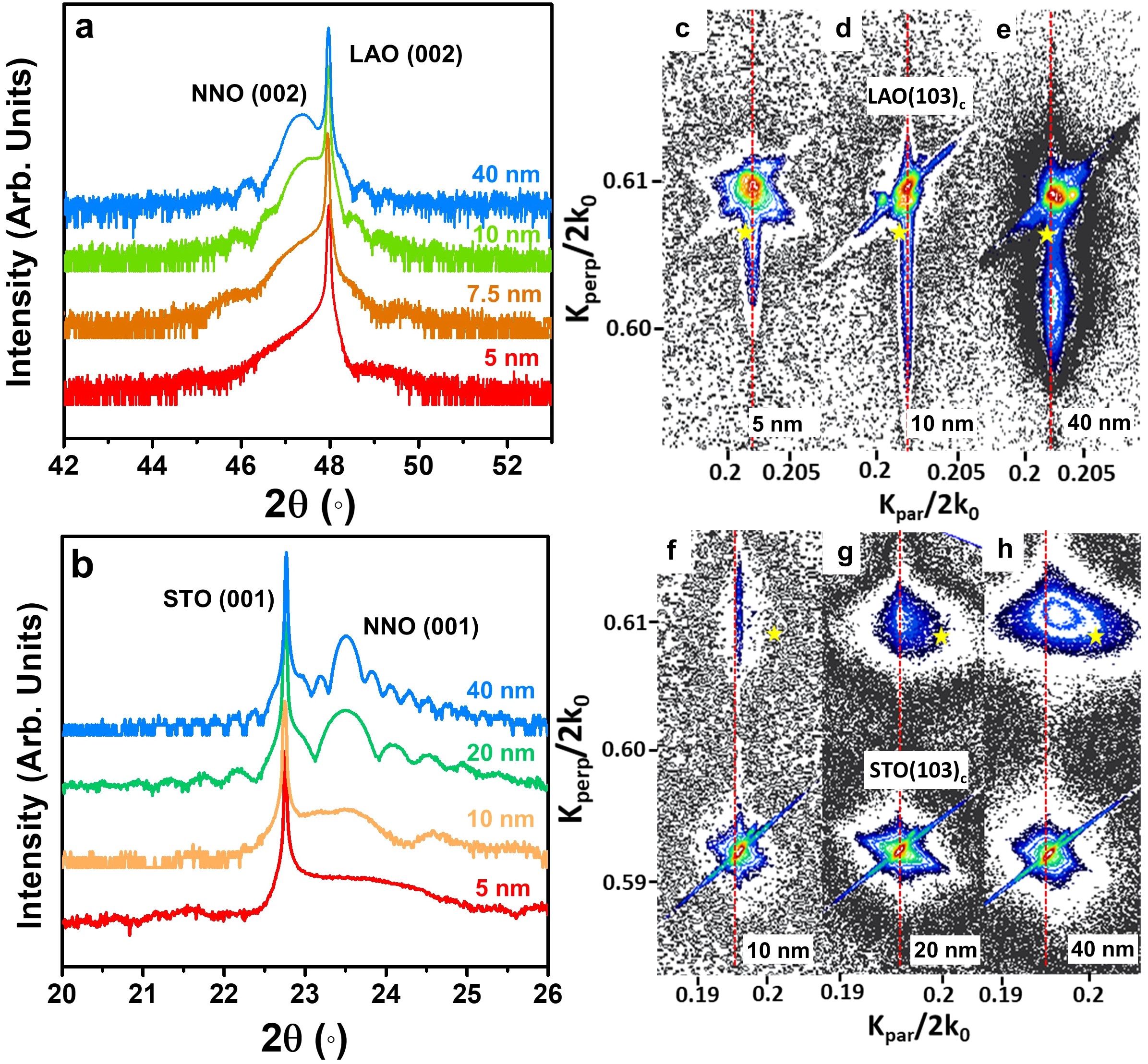}
\caption{X-ray diffraction patterns around the 002-reflection of NNO/LAO films (a) and around the 001-reflection of NNO/STO films (b) with different thicknesses. Reciprocal space map (RSM) around the (103)\textsubscript{c} diffraction peaks of (c) 5 nm, (d) 10 nm, (e) 40 nm NNO/LAO films and (f) 10 nm, (g) 20 nm, (h) 40 nm NNO/STO films. The red dashed lines are guides to the eyes showing the substrate in-plane lattice. The yellow stars signals the (103)\textsubscript{c} peak of bulk NNO.}
\label{fig:XRD}
\end{figure}

Figures \ref{fig:XRD}(a)-(b) show the diffraction patterns for films grown on LAO and STO, respectively. The presence of Laue fringes indicates the high quality of the interfaces. The different sign of the epitaxial strain on the two substrates can be assessed by the different relative positions of the film and substrate peaks. Reciprocal space maps (RSM) around the (103)\textsubscript{c} peaks, are shown in Fig. \ref{fig:XRD}(c-h). All NNO films grown on LAO grow coherently with the substrate (with coincident in-plane reciprocal lattices of film and substrate), for all investigated thicknesses, as expected from the very similar lattice of the bulk NNO (signalled in the maps by the yellow stars) and the substrate. On the contrary, in the NNO/STO films, only the thinnest films grow coherently with the substrate and show an in-plane lattice significantly larger than that of the bulk, due to the large differences between the bulk NNO and the STO substrate lattices. For increasing thicknesses, a gradual shift of the film peak can be observed, in agreement with the expected evolution of the lattice parameters and strain relaxation toward the bulk lattice, with increasing thickness. Thus, the observed evolution of \textit{n} (Fig. \ref{fig:fitparameters}(a)) corresponds to the gradually relaxed in-plane strain of the films. 

Figure \ref{fig:fitparameters}(b) summarizes the \textit{n} values extracted from the NNO films as a function of the in-plane strain, $\varepsilon$\textsubscript{xx}, obtained from the diffraction data in Fig.  \ref{fig:XRD}. Data from NNO films on NGO substrates ($\varepsilon$\textsubscript{xx}= +1.34\%) are also included. A 5 nm NNO/NGO film also shows apparent linear \textit{T} scaling in the metallic phase, confirming the correlation between the magnitude of the tensile strain and \textit{n} (see Fig. S4 in SM). The extended resistivity data (inset of Fig. S4) also shows a non-hysteretic phase transition similar to that of NNO/STO. Figure \ref{fig:fitparameters}(b) is completed with \textit{n} values reported by other authors for bulk NNO \cite{blasco1994structural} and NNO films under larger compressive strains \cite{mikheev2015tuning, liu2013heterointerface}. Indeed, we observe a clear dependence of \textit{n} on the in-plane strain. Both tensile and compressive strains are expected to induce an increase of the orbital splitting between the Ni$^{3+}$ ${x^2-y^2}$ and ${3z^2-r^2}$ $e_g$ levels \cite{mikheev2015tuning}. However, the large asymmetry observed, with a significantly stronger dependence for the tensile strain regime, points to an additional influence on \textit{n}.

\begin{figure}
\centering
\includegraphics[width=.43\textwidth]{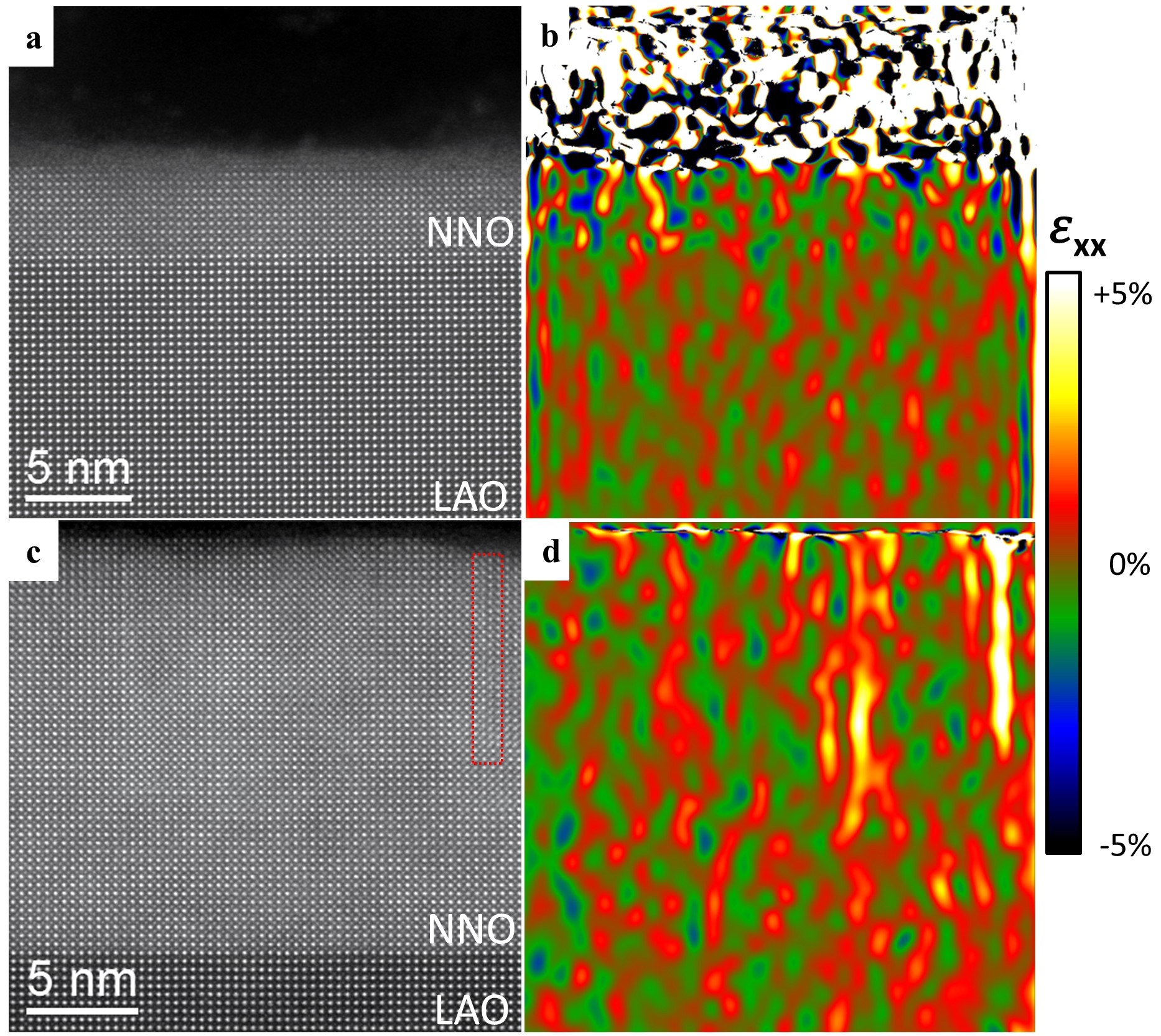}
\caption{Cross-sectional HAADF-STEM image of NdNiO\textsubscript{3} thin films grown on a LaAlO\textsubscript{3} substrates, for a 5 nm thick film (a) and a 20 nm thick film (c). The respective in-plane components of the strain tensor obtained from the STEM images by geometrical phase analysis (GPA) are shown in (b) and (d). The red dashed lines surround the RP faults.}
\label{fig:TEM1}
\end{figure}

\begin{figure}
\centering
\includegraphics[width=.45\textwidth]{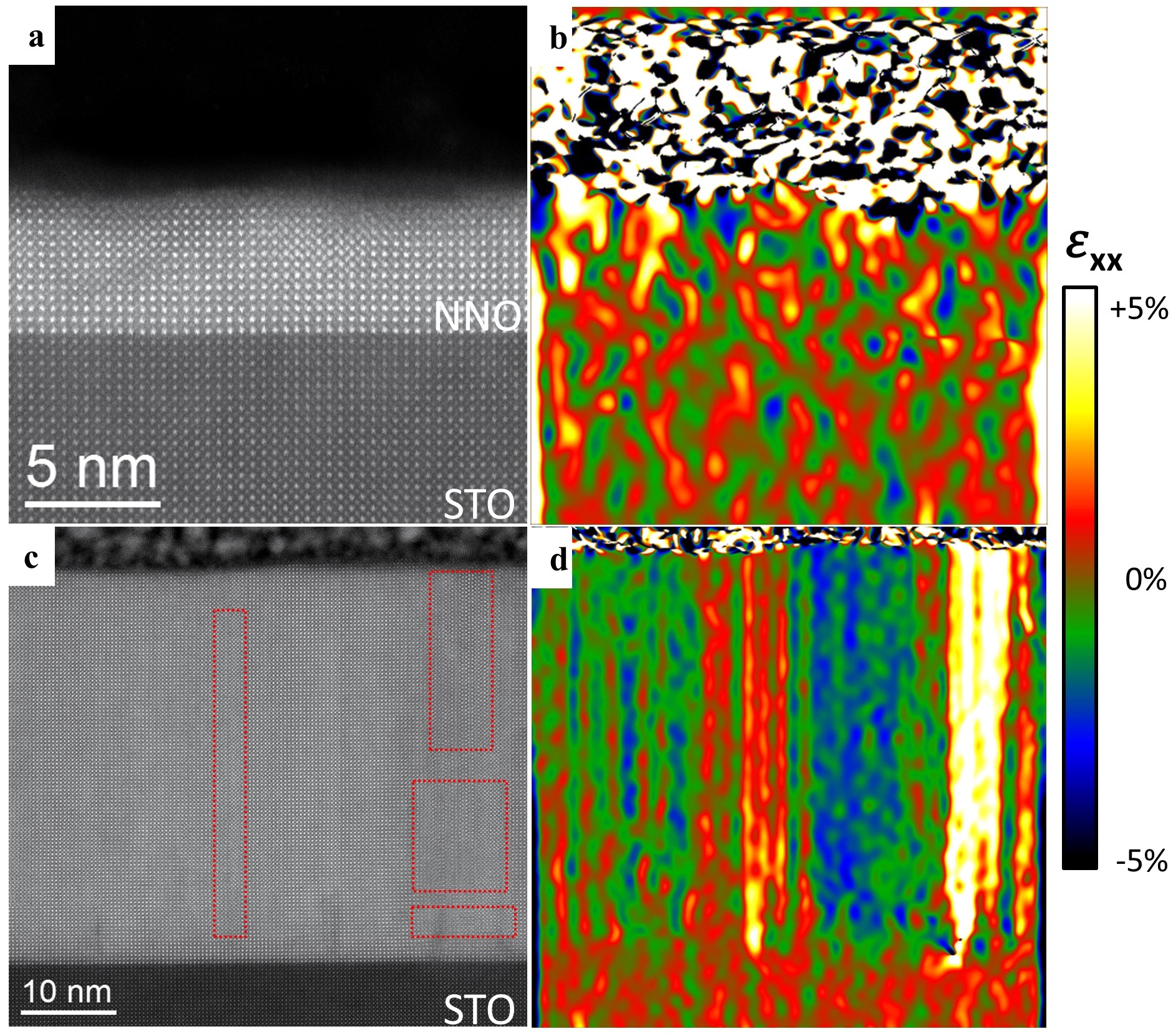}
\caption{Cross-sectional HAADF-STEM image of NdNiO\textsubscript{3} thin films grown on a SrTiO\textsubscript{3} substrates, for a 5 nm thick film (a) and a 40 nm thick film (c). The respective in-plane components of the strain tensor obtained from the STEM images by geometrical phase analysis (GPA) are shown in (b) and (d). The red dashed lines surround the RP faults.}
\label{fig:TEM2}
\end{figure}

In order to shed light into this behaviour, we performed scanning transmission electron microscopy (STEM) on the films. Cross-sectional specimens of the films were studied by atomic resolution STEM (for experimental details, see SM). The high-angle annular dark-field (HAADF) STEM image shown in Fig. \ref{fig:TEM1}(a) evidences the epitaxial, cube-on-cube growth of a 5 nm thick NNO film on a LAO substrate, with a flat, atomically sharp interface. No defects or misfit dislocations are observed. The strain state of the films was determined by geometrical phase analysis (GPA) of the HAADF images, the deformation of the in-plane lattice parameter of the film respect to the substrate ($\varepsilon_{xx}$) is depicted in Fig. \ref{fig:TEM1}(b). $\varepsilon_{xx}$ is virtually zero across the 5 nm NNO film, showing a good in-plane lattice match between film and substrate, in agreement with the x-ray diffraction data. The 20 nm thick film on LAO substrate also shows $\varepsilon_{xx}$= 0 across most of the film but it starts showing small regions with  Ruddlesden-Popper (RP) faults, often reported in nickelates \cite{bak2017formation}, as seen in Fig. \ref{fig:TEM1}(c)-(d). Some effect of these RP defects can be seen in the electrical properties, which show a strongly decreased resistance in the insulating state (Fig. \ref{fig:resistivity}(a)), as well as a increased resistivity in the metallic state for the 40 nm films on LAO (Fig. \ref{fig:resistivity}(c)). However, the PR defects do not preclude the presence of hysteresis at the metal-insulator transition, nor the apparent linear behaviour of the metallic resistivity in Fig. \ref{fig:resistivity}(c). RP faults are known to have a significantly enlarged out-of-plane lattice parameter \cite{bak2017formation}, which can explain the unusual evolution of the out-of-plane lattice parameters as a function of thickness for the NNO/LAO films, shown in Fig. \ref{fig:XRD}(a). Similar images for the thinnest and the thickest films on STO, shown in Figure \ref{fig:TEM2}, reveal a higher abundance of RP faults, which are present even in the thinnest films. The data, thus, strongly suggest that the RP secondary phases present in the films, are not correlated with the observed changes of \textit{n}.
 
\begin{figure}
\includegraphics[width=0.35\textwidth]{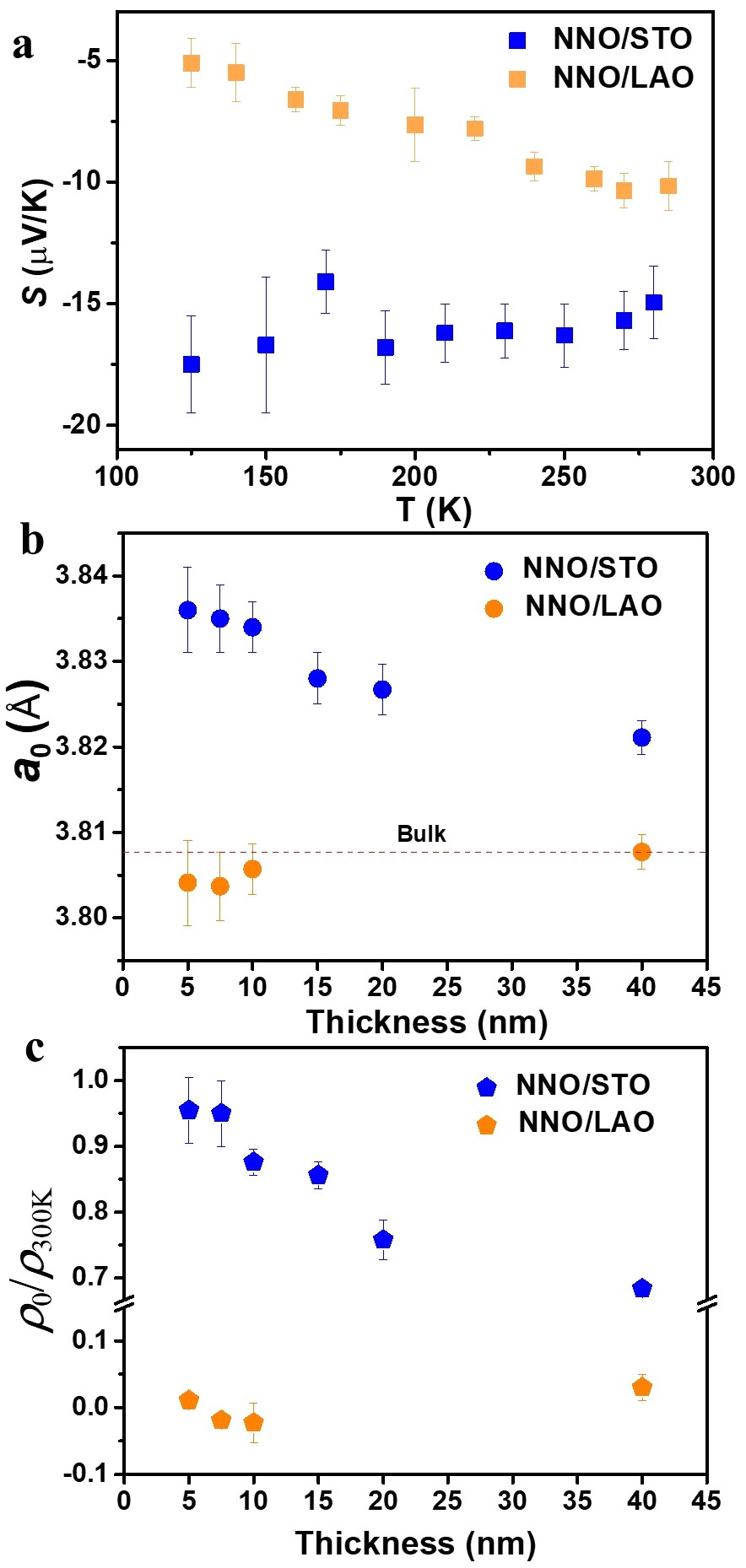}
\caption{(a) Seebeck coefficients measured on NNO films with thickness on 10 nm grown on LAO and STO substrates. (b) The unstrained film lattice parameter (a\textsubscript{0}) and (c) the reversed residual-resistivity ratio ($\rho$\textsubscript{0}/$\rho$\textsubscript{300K}) of NNO films grown on LAO and STO substrates with different thickness.}
\label{fig:Seebeck}
\end{figure}

 The effect of strain on \textit{n} may be indirect. Planar defects such as misfit dislocations or stacking faults have been often observed in nickelate films \cite{coll2017simulation} and the creation of oxygen vacancies is known to be an efficient mechanism to relax tensile strain in epitaxially grown perovskites, as oxygen vacancies locally enlarge the lattice \cite{catalano2018rare,catalan2008progress,lopez2017evidence}. In nickelate thin films, a pair of oxygen vacancies favour the reduction of the Ni ions to Ni\textsuperscript{2+} \cite{malashevich2015first,shi2013correlated,wang2016oxygen}. Indeed, measurement of Seebeck coefficients on films with a thickness of 10 nm grown on LAO and STO, plotted in Figure \ref{fig:Seebeck}(a), show that while the film on LAO displays metallic-like transport, the film on STO shows a flat temperature dependence, characteristics of polaronic systems.

 Another indication of the existence of an increased content of oxygen vacancies in our films on STO comes from the structural data. From the definition of Poisson ratio, $\nu $, the pseudo-cubic lattice parameters that would correspond to the unstrained case for the different films can be estimated as $a_o$ = (2$\nu$a+(1-$\nu$)c)/(1+$\nu$) \cite{hauser2015correlation, iglesias2017oxygen}, where $a$ and $c$ are the in-plane and out-of-plane lattice parameters of the films, respectively, obtained from the structural data of Fig. \ref{fig:XRD}, and $\nu$= 0.30 has been used for all films. The results, plotted in Figure \ref{fig:Seebeck}(b), show that the films on LAO display a lattice volume close to the bulk value, while the unit cell volume of the films on STO is significantly increased, which is consistent with a larger oxygen vacancy content that decreases with increasing thickness. Moreover, the residual resistivity ratio (RRR), which is often used as a measurement of materials purity, increases with increasing thickness in the films on STO, also in agreement with a lower vacancy content in the thicker films.  

Our experiments, therefore, indicate that NNO films subjected to relatively small strain values, display \textit{T}-linear resistivity scaling. For larger values of tensile strain, an increase of the power law resistivity-temperature exponent with the magnitude of the strain is observed. This is related to both the effect of strain on the orbital splitting and the degree of disorder, most likely due to oxygen vacancies, whose concentration is believed to increase with increasing tensile strain. These results validate recent theoretical predictions by Patel \textit{et al.} \cite{patel2017non}. Their computational work uses the Anderson-Hubbard Hamiltonian to predict that the metallic state that arises for small and intermediate values of both the on-site Coulomb interaction of 3$d$-electrons (\textit{U}) and the disorder (\textit{V}) can be continuously tuned. The calculations predict values varying from \textit{n}= 1 to \textit{n}= 2 by the joint action of both \textit{U} and \textit{V}; while in our experiments, larger values up to \textit{n}= 3 are also observed. Power law exponents varying with the degree of disorder have also been reported for SrRuO$_3$ thin films by Herranz et al. \cite{herranz2008effect}.

Accordingly, large enough tensile strain should induce a large density of oxygen vacancies and should, eventually, suppress the metallic phase. This is confirmed in films grown on DSO substrates, under +3.86 \% strain, for which the resistivity data can be described by a variable range hopping (VRH) conduction model for T$<$ 70 K (see Figure S5 in SM) followed by a nearest neighbour hopping (NNH) model with E$_a$= 32 meV for temperatures above T= 70 K, as often observed in disordered solids \cite{mott1969conduction, issai2015hopping}. It is interesting to notice that a film of the same thickness on STO show similar behaviour: comparable E$_a$ in the NNH regime and comparable crossover temperature from VRH to NNH conduction (see Figure S6 in SM). It is known that the presence of quenched disorder strongly impacts the transport properties inducing percolation and changing the nature of the phase transition \cite{Salamon2002Colossal}. In such percolation picture, a coexistence of metallic and insulating clusters could persist into the metallic phase. Indeed, the data of the films under intermediate strain (on STO) shows a magnitude of the resistivity in the metallic state that is in between those of the film on DSO and the film on LAO. It is worth to point out that oxygen vacancies can also order in nickelates, as recently shown both in thin films \cite{coll2017simulation} and bulk crystals \cite{wang2018antiferromagnetic} of metallic LaNiO$_{3-\delta}$. The controlled tunability of oxygen vacancies with strain and its direct relationship with the transport properties demonstrated could also be of importance in the context of the bond disproportionation and negative charge transfer models \cite{bisogni2016ground}, as well as the recent work proposing the metal state as a bi-polaron liquid and the insulating phase as its ordered (bond-disproportionated) version \cite{shamblin2018experimental}.

To summarize, this work reports a clear evolution of the apparent scaling exponent of the resistivity-temperature characteristics (\textit{n}) with strain and disorder, supporting recent theoretical predictions that show the tunability of the scaling exponents arising from the interplay between electron interactions and disorder in nickelates \cite{patel2017non}. The overall picture helps to clarify that the underlying physics behind the observed evolution of exponents from \textit{T}-linear to quadratic scaling and beyond, does not necessarily imply a crossover between FL and NFL behaviour or other exotic physics. On the contrary, for the films reported here with bulk-like in-plane lattice parameters, the contribution to the transport properties from delocalized electrons for the intermediate temperature region above the metal-insulator transition is fully consistent with a classical Fermi gas ruled by electron-phonon scattering.

\section{Acknowledgements}

We are grateful to Manuel Bibes, Nigel Hussey and Jan Zaanen for insightful discussions that have helped reaching the current form of the manuscript. We also thankfully acknowledge useful discussions with Graeme Blake, Erik van Heumen, Pavan Nukala, Mart Salverda and Arjun Joshua. In addition, We want to thank Jacob Baas and Henk Bonder for their invaluable technical support. Qikai Guo and Saeedeh Farokhipoor acknowledge financial support from a China Scholarship Council (CSC) grant and a VENI grant (016.veni.179.053) of the Netherlands Organisation for Scientific Research (NWO), respectively. F.R. acknowledges support by the Ministry of Science of Spain (Project No. MAT2016-80762-R), the Conselleria de Cultura, Educacion e Ordenacion Universitaria (ED431F 2016/008, and Centro singular de investigación de Galicia accreditation 2016-2019, ED431G/09), the Xunta de Galicia, the European Social Fund (ESF) and the European Regional Development Fund (ERDF).

\bibliography{Reference}

\begin{thebibliography}{62}%
\makeatletter
\providecommand \@ifxundefined [1]{%
 \@ifx{#1\undefined}
}%
\providecommand \@ifnum [1]{%
 \ifnum #1\expandafter \@firstoftwo
 \else \expandafter \@secondoftwo
 \fi
}%
\providecommand \@ifx [1]{%
 \ifx #1\expandafter \@firstoftwo
 \else \expandafter \@secondoftwo
 \fi
}%
\providecommand \natexlab [1]{#1}%
\providecommand \enquote  [1]{``#1''}%
\providecommand \bibnamefont  [1]{#1}%
\providecommand \bibfnamefont [1]{#1}%
\providecommand \citenamefont [1]{#1}%
\providecommand \href@noop [0]{\@secondoftwo}%
\providecommand \href [0]{\begingroup \@sanitize@url \@href}%
\providecommand \@href[1]{\@@startlink{#1}\@@href}%
\providecommand \@@href[1]{\endgroup#1\@@endlink}%
\providecommand \@sanitize@url [0]{\catcode `\\12\catcode `\$12\catcode
  `\&12\catcode `\#12\catcode `\^12\catcode `\_12\catcode `\%12\relax}%
\providecommand \@@startlink[1]{}%
\providecommand \@@endlink[0]{}%
\providecommand \url  [0]{\begingroup\@sanitize@url \@url }%
\providecommand \@url [1]{\endgroup\@href {#1}{\urlprefix }}%
\providecommand \urlprefix  [0]{URL }%
\providecommand \Eprint [0]{\href }%
\providecommand \doibase [0]{http://dx.doi.org/}%
\providecommand \selectlanguage [0]{\@gobble}%
\providecommand \bibinfo  [0]{\@secondoftwo}%
\providecommand \bibfield  [0]{\@secondoftwo}%
\providecommand \translation [1]{[#1]}%
\providecommand \BibitemOpen [0]{}%
\providecommand \bibitemStop [0]{}%
\providecommand \bibitemNoStop [0]{.\EOS\space}%
\providecommand \EOS [0]{\spacefactor3000\relax}%
\providecommand \BibitemShut  [1]{\csname bibitem#1\endcsname}%
\let\auto@bib@innerbib\@empty
\bibitem [{\citenamefont {Scherwitzl}\ \emph {et~al.}(2010)\citenamefont
  {Scherwitzl}, \citenamefont {Zubko}, \citenamefont {Lezama}, \citenamefont
  {Ono}, \citenamefont {Morpurgo}, \citenamefont {Catalan},\ and\ \citenamefont
  {Triscone}}]{scherwitzl2010electric}%
  \BibitemOpen
  \bibfield  {author} {\bibinfo {author} {\bibfnamefont {R.}~\bibnamefont
  {Scherwitzl}}, \bibinfo {author} {\bibfnamefont {P.}~\bibnamefont {Zubko}},
  \bibinfo {author} {\bibfnamefont {I.~G.}\ \bibnamefont {Lezama}}, \bibinfo
  {author} {\bibfnamefont {S.}~\bibnamefont {Ono}}, \bibinfo {author}
  {\bibfnamefont {A.~F.}\ \bibnamefont {Morpurgo}}, \bibinfo {author}
  {\bibfnamefont {G.}~\bibnamefont {Catalan}}, \ and\ \bibinfo {author}
  {\bibfnamefont {J.-M.}\ \bibnamefont {Triscone}},\ }\href@noop {} {\bibfield
  {journal} {\bibinfo  {journal} {Advanced Materials}\ }\textbf {\bibinfo
  {volume} {22}},\ \bibinfo {pages} {5517} (\bibinfo {year}
  {2010})}\BibitemShut {NoStop}%
\bibitem [{\citenamefont {Driscoll}\ \emph {et~al.}(2009)\citenamefont
  {Driscoll}, \citenamefont {Kim}, \citenamefont {Chae}, \citenamefont
  {Di~Ventra},\ and\ \citenamefont {Basov}}]{driscoll2009phase}%
  \BibitemOpen
  \bibfield  {author} {\bibinfo {author} {\bibfnamefont {T.}~\bibnamefont
  {Driscoll}}, \bibinfo {author} {\bibfnamefont {H.-T.}\ \bibnamefont {Kim}},
  \bibinfo {author} {\bibfnamefont {B.-G.}\ \bibnamefont {Chae}}, \bibinfo
  {author} {\bibfnamefont {M.}~\bibnamefont {Di~Ventra}}, \ and\ \bibinfo
  {author} {\bibfnamefont {D.}~\bibnamefont {Basov}},\ }\href@noop {}
  {\bibfield  {journal} {\bibinfo  {journal} {Applied physics letters}\
  }\textbf {\bibinfo {volume} {95}},\ \bibinfo {pages} {043503} (\bibinfo
  {year} {2009})}\BibitemShut {NoStop}%
\bibitem [{\citenamefont {McLeod}\ \emph {et~al.}(2017)\citenamefont {McLeod},
  \citenamefont {Van~Heumen}, \citenamefont {Ramirez}, \citenamefont {Wang},
  \citenamefont {Saerbeck}, \citenamefont {Guenon}, \citenamefont {Goldflam},
  \citenamefont {Anderegg}, \citenamefont {Kelly}, \citenamefont {Mueller}
  \emph {et~al.}}]{mcleod2017nanotextured}%
  \BibitemOpen
  \bibfield  {author} {\bibinfo {author} {\bibfnamefont {A.}~\bibnamefont
  {McLeod}}, \bibinfo {author} {\bibfnamefont {E.}~\bibnamefont {Van~Heumen}},
  \bibinfo {author} {\bibfnamefont {J.}~\bibnamefont {Ramirez}}, \bibinfo
  {author} {\bibfnamefont {S.}~\bibnamefont {Wang}}, \bibinfo {author}
  {\bibfnamefont {T.}~\bibnamefont {Saerbeck}}, \bibinfo {author}
  {\bibfnamefont {S.}~\bibnamefont {Guenon}}, \bibinfo {author} {\bibfnamefont
  {M.}~\bibnamefont {Goldflam}}, \bibinfo {author} {\bibfnamefont
  {L.}~\bibnamefont {Anderegg}}, \bibinfo {author} {\bibfnamefont
  {P.}~\bibnamefont {Kelly}}, \bibinfo {author} {\bibfnamefont
  {A.}~\bibnamefont {Mueller}},  \emph {et~al.},\ }\href@noop {} {\bibfield
  {journal} {\bibinfo  {journal} {Nature Physics}\ }\textbf {\bibinfo {volume}
  {13}},\ \bibinfo {pages} {80} (\bibinfo {year} {2017})}\BibitemShut {NoStop}%
\bibitem [{\citenamefont {Ha}\ \emph {et~al.}(2011)\citenamefont {Ha},
  \citenamefont {Aydogdu},\ and\ \citenamefont {Ramanathan}}]{ha2011metal}%
  \BibitemOpen
  \bibfield  {author} {\bibinfo {author} {\bibfnamefont {S.~D.}\ \bibnamefont
  {Ha}}, \bibinfo {author} {\bibfnamefont {G.~H.}\ \bibnamefont {Aydogdu}}, \
  and\ \bibinfo {author} {\bibfnamefont {S.}~\bibnamefont {Ramanathan}},\
  }\href@noop {} {\bibfield  {journal} {\bibinfo  {journal} {Applied Physics
  Letters}\ }\textbf {\bibinfo {volume} {98}},\ \bibinfo {pages} {012105}
  (\bibinfo {year} {2011})}\BibitemShut {NoStop}%
\bibitem [{\citenamefont {Ha}\ \emph {et~al.}(2014)\citenamefont {Ha},
  \citenamefont {Shi}, \citenamefont {Meroz}, \citenamefont {Mahadevan},\ and\
  \citenamefont {Ramanathan}}]{ha2014neuromimetic}%
  \BibitemOpen
  \bibfield  {author} {\bibinfo {author} {\bibfnamefont {S.~D.}\ \bibnamefont
  {Ha}}, \bibinfo {author} {\bibfnamefont {J.}~\bibnamefont {Shi}}, \bibinfo
  {author} {\bibfnamefont {Y.}~\bibnamefont {Meroz}}, \bibinfo {author}
  {\bibfnamefont {L.}~\bibnamefont {Mahadevan}}, \ and\ \bibinfo {author}
  {\bibfnamefont {S.}~\bibnamefont {Ramanathan}},\ }\href@noop {} {\bibfield
  {journal} {\bibinfo  {journal} {Physical Review Applied}\ }\textbf {\bibinfo
  {volume} {2}},\ \bibinfo {pages} {064003} (\bibinfo {year}
  {2014})}\BibitemShut {NoStop}%
\bibitem [{\citenamefont {Shi}\ \emph {et~al.}(2013)\citenamefont {Shi},
  \citenamefont {Ha}, \citenamefont {Zhou}, \citenamefont {Schoofs},\ and\
  \citenamefont {Ramanathan}}]{shi2013correlated}%
  \BibitemOpen
  \bibfield  {author} {\bibinfo {author} {\bibfnamefont {J.}~\bibnamefont
  {Shi}}, \bibinfo {author} {\bibfnamefont {S.~D.}\ \bibnamefont {Ha}},
  \bibinfo {author} {\bibfnamefont {Y.}~\bibnamefont {Zhou}}, \bibinfo {author}
  {\bibfnamefont {F.}~\bibnamefont {Schoofs}}, \ and\ \bibinfo {author}
  {\bibfnamefont {S.}~\bibnamefont {Ramanathan}},\ }\href@noop {} {\bibfield
  {journal} {\bibinfo  {journal} {Nature communications}\ }\textbf {\bibinfo
  {volume} {4}},\ \bibinfo {pages} {2676} (\bibinfo {year} {2013})}\BibitemShut
  {NoStop}%
\bibitem [{\citenamefont {Wang}\ \emph {et~al.}(2017)\citenamefont {Wang},
  \citenamefont {Zhang}, \citenamefont {Chang}, \citenamefont {You},
  \citenamefont {He}, \citenamefont {Jin}, \citenamefont {Gu}, \citenamefont
  {Guo}, \citenamefont {Ge}, \citenamefont {Feng} \emph
  {et~al.}}]{wang2017electrochemically}%
  \BibitemOpen
  \bibfield  {author} {\bibinfo {author} {\bibfnamefont {L.}~\bibnamefont
  {Wang}}, \bibinfo {author} {\bibfnamefont {Q.}~\bibnamefont {Zhang}},
  \bibinfo {author} {\bibfnamefont {L.}~\bibnamefont {Chang}}, \bibinfo
  {author} {\bibfnamefont {L.}~\bibnamefont {You}}, \bibinfo {author}
  {\bibfnamefont {X.}~\bibnamefont {He}}, \bibinfo {author} {\bibfnamefont
  {K.}~\bibnamefont {Jin}}, \bibinfo {author} {\bibfnamefont {L.}~\bibnamefont
  {Gu}}, \bibinfo {author} {\bibfnamefont {H.}~\bibnamefont {Guo}}, \bibinfo
  {author} {\bibfnamefont {C.}~\bibnamefont {Ge}}, \bibinfo {author}
  {\bibfnamefont {Y.}~\bibnamefont {Feng}},  \emph {et~al.},\ }\href@noop {}
  {\bibfield  {journal} {\bibinfo  {journal} {Advanced Electronic Materials}\
  }\textbf {\bibinfo {volume} {3}},\ \bibinfo {pages} {1700321} (\bibinfo
  {year} {2017})}\BibitemShut {NoStop}%
\bibitem [{\citenamefont {Emery}\ and\ \citenamefont
  {Kivelson}(1995)}]{emery1995superconductivity}%
  \BibitemOpen
  \bibfield  {author} {\bibinfo {author} {\bibfnamefont {V.}~\bibnamefont
  {Emery}}\ and\ \bibinfo {author} {\bibfnamefont {S.}~\bibnamefont
  {Kivelson}},\ }\href@noop {} {\bibfield  {journal} {\bibinfo  {journal}
  {Physical Review Letters}\ }\textbf {\bibinfo {volume} {74}},\ \bibinfo
  {pages} {3253} (\bibinfo {year} {1995})}\BibitemShut {NoStop}%
\bibitem [{\citenamefont {Zaanen}(2019)}]{zaanen2019}%
  \BibitemOpen
  \bibfield  {author} {\bibinfo {author} {\bibfnamefont {J.}~\bibnamefont
  {Zaanen}},\ }\href {\doibase 10.21468/SciPostPhys.6.5.061} {\bibfield
  {journal} {\bibinfo  {journal} {SciPost Phys.}\ }\textbf {\bibinfo {volume}
  {6}},\ \bibinfo {pages} {61} (\bibinfo {year} {2019})}\BibitemShut {NoStop}%
\bibitem [{\citenamefont {Landau}\ \emph {et~al.}(1956)\citenamefont {Landau},
  \citenamefont {Abrikosov},\ and\ \citenamefont
  {Khalatnikov}}]{landau1956multiplicative}%
  \BibitemOpen
  \bibfield  {author} {\bibinfo {author} {\bibfnamefont {L.}~\bibnamefont
  {Landau}}, \bibinfo {author} {\bibfnamefont {A.}~\bibnamefont {Abrikosov}}, \
  and\ \bibinfo {author} {\bibfnamefont {I.}~\bibnamefont {Khalatnikov}},\
  }\href@noop {} {\bibfield  {journal} {\bibinfo  {journal} {Soviet Physics
  JETP}\ }\textbf {\bibinfo {volume} {3}} (\bibinfo {year} {1956})}\BibitemShut
  {NoStop}%
\bibitem [{\citenamefont {Stewart}(2001)}]{stewart2001non}%
  \BibitemOpen
  \bibfield  {author} {\bibinfo {author} {\bibfnamefont {G.}~\bibnamefont
  {Stewart}},\ }\href@noop {} {\bibfield  {journal} {\bibinfo  {journal}
  {Reviews of modern Physics}\ }\textbf {\bibinfo {volume} {73}},\ \bibinfo
  {pages} {797} (\bibinfo {year} {2001})}\BibitemShut {NoStop}%
\bibitem [{\citenamefont {Schofield}(1999)}]{schofield1999non}%
  \BibitemOpen
  \bibfield  {author} {\bibinfo {author} {\bibfnamefont {A.~J.}\ \bibnamefont
  {Schofield}},\ }\href@noop {} {\bibfield  {journal} {\bibinfo  {journal}
  {Contemporary Physics}\ }\textbf {\bibinfo {volume} {40}},\ \bibinfo {pages}
  {95} (\bibinfo {year} {1999})}\BibitemShut {NoStop}%
\bibitem [{\citenamefont {Rivadulla}\ \emph {et~al.}(2007)\citenamefont
  {Rivadulla}, \citenamefont {Fern{\'a}ndez-Rossier}, \citenamefont
  {Garc{\'\i}a-Hern{\'a}ndez}, \citenamefont {L{\'o}pez-Quintela},
  \citenamefont {Rivas},\ and\ \citenamefont {Goodenough}}]{rivadulla2007vo}%
  \BibitemOpen
  \bibfield  {author} {\bibinfo {author} {\bibfnamefont {F.}~\bibnamefont
  {Rivadulla}}, \bibinfo {author} {\bibfnamefont {J.}~\bibnamefont
  {Fern{\'a}ndez-Rossier}}, \bibinfo {author} {\bibfnamefont {M.}~\bibnamefont
  {Garc{\'\i}a-Hern{\'a}ndez}}, \bibinfo {author} {\bibfnamefont
  {M.}~\bibnamefont {L{\'o}pez-Quintela}}, \bibinfo {author} {\bibfnamefont
  {J.}~\bibnamefont {Rivas}}, \ and\ \bibinfo {author} {\bibfnamefont
  {J.}~\bibnamefont {Goodenough}},\ }\href@noop {} {\bibfield  {journal}
  {\bibinfo  {journal} {Physical Review B}\ }\textbf {\bibinfo {volume} {76}},\
  \bibinfo {pages} {205110} (\bibinfo {year} {2007})}\BibitemShut {NoStop}%
\bibitem [{\citenamefont {Stemmer}\ and\ \citenamefont
  {Allen}(2018)}]{stemmer2018non}%
  \BibitemOpen
  \bibfield  {author} {\bibinfo {author} {\bibfnamefont {S.}~\bibnamefont
  {Stemmer}}\ and\ \bibinfo {author} {\bibfnamefont {S.~J.}\ \bibnamefont
  {Allen}},\ }\href@noop {} {\bibfield  {journal} {\bibinfo  {journal} {Reports
  on Progress in Physics}\ }\textbf {\bibinfo {volume} {81}},\ \bibinfo {pages}
  {062502} (\bibinfo {year} {2018})}\BibitemShut {NoStop}%
\bibitem [{\citenamefont {Keller}\ \emph {et~al.}(2015)\citenamefont {Keller},
  \citenamefont {Peeters}, \citenamefont {Moca}, \citenamefont {Weymann},
  \citenamefont {Mahalu}, \citenamefont {Umansky}, \citenamefont {Zar{\'a}nd},\
  and\ \citenamefont {Goldhaber-Gordon}}]{keller2015universal}%
  \BibitemOpen
  \bibfield  {author} {\bibinfo {author} {\bibfnamefont {A.}~\bibnamefont
  {Keller}}, \bibinfo {author} {\bibfnamefont {L.}~\bibnamefont {Peeters}},
  \bibinfo {author} {\bibfnamefont {C.}~\bibnamefont {Moca}}, \bibinfo {author}
  {\bibfnamefont {I.}~\bibnamefont {Weymann}}, \bibinfo {author} {\bibfnamefont
  {D.}~\bibnamefont {Mahalu}}, \bibinfo {author} {\bibfnamefont
  {V.}~\bibnamefont {Umansky}}, \bibinfo {author} {\bibfnamefont
  {G.}~\bibnamefont {Zar{\'a}nd}}, \ and\ \bibinfo {author} {\bibfnamefont
  {D.}~\bibnamefont {Goldhaber-Gordon}},\ }\href@noop {} {\bibfield  {journal}
  {\bibinfo  {journal} {Nature}\ }\textbf {\bibinfo {volume} {526}},\ \bibinfo
  {pages} {237} (\bibinfo {year} {2015})}\BibitemShut {NoStop}%
\bibitem [{\citenamefont {Kasahara}\ \emph {et~al.}(2010)\citenamefont
  {Kasahara}, \citenamefont {Shibauchi}, \citenamefont {Hashimoto},
  \citenamefont {Ikada}, \citenamefont {Tonegawa}, \citenamefont {Okazaki},
  \citenamefont {Shishido}, \citenamefont {Ikeda}, \citenamefont {Takeya},
  \citenamefont {Hirata} \emph {et~al.}}]{kasahara2010evolution}%
  \BibitemOpen
  \bibfield  {author} {\bibinfo {author} {\bibfnamefont {S.}~\bibnamefont
  {Kasahara}}, \bibinfo {author} {\bibfnamefont {T.}~\bibnamefont {Shibauchi}},
  \bibinfo {author} {\bibfnamefont {K.}~\bibnamefont {Hashimoto}}, \bibinfo
  {author} {\bibfnamefont {K.}~\bibnamefont {Ikada}}, \bibinfo {author}
  {\bibfnamefont {S.}~\bibnamefont {Tonegawa}}, \bibinfo {author}
  {\bibfnamefont {R.}~\bibnamefont {Okazaki}}, \bibinfo {author} {\bibfnamefont
  {H.}~\bibnamefont {Shishido}}, \bibinfo {author} {\bibfnamefont
  {H.}~\bibnamefont {Ikeda}}, \bibinfo {author} {\bibfnamefont
  {H.}~\bibnamefont {Takeya}}, \bibinfo {author} {\bibfnamefont
  {K.}~\bibnamefont {Hirata}},  \emph {et~al.},\ }\href@noop {} {\bibfield
  {journal} {\bibinfo  {journal} {Physical Review B}\ }\textbf {\bibinfo
  {volume} {81}},\ \bibinfo {pages} {184519} (\bibinfo {year}
  {2010})}\BibitemShut {NoStop}%
\bibitem [{\citenamefont {Lee}(2018)}]{lee2018recent}%
  \BibitemOpen
  \bibfield  {author} {\bibinfo {author} {\bibfnamefont {S.-S.}\ \bibnamefont
  {Lee}},\ }\href@noop {} {\bibfield  {journal} {\bibinfo  {journal} {Annual
  Review of Condensed Matter Physics}\ }\textbf {\bibinfo {volume} {9}},\
  \bibinfo {pages} {227} (\bibinfo {year} {2018})}\BibitemShut {NoStop}%
\bibitem [{\citenamefont {Imada}\ \emph {et~al.}(1998)\citenamefont {Imada},
  \citenamefont {Fujimori},\ and\ \citenamefont {Tokura}}]{imada1998metal}%
  \BibitemOpen
  \bibfield  {author} {\bibinfo {author} {\bibfnamefont {M.}~\bibnamefont
  {Imada}}, \bibinfo {author} {\bibfnamefont {A.}~\bibnamefont {Fujimori}}, \
  and\ \bibinfo {author} {\bibfnamefont {Y.}~\bibnamefont {Tokura}},\
  }\href@noop {} {\bibfield  {journal} {\bibinfo  {journal} {Reviews of modern
  physics}\ }\textbf {\bibinfo {volume} {70}},\ \bibinfo {pages} {1039}
  (\bibinfo {year} {1998})}\BibitemShut {NoStop}%
\bibitem [{\citenamefont {Medarde}(1997)}]{medarde1997structural}%
  \BibitemOpen
  \bibfield  {author} {\bibinfo {author} {\bibfnamefont {M.~L.}\ \bibnamefont
  {Medarde}},\ }\href@noop {} {\bibfield  {journal} {\bibinfo  {journal}
  {Journal of Physics: Condensed Matter}\ }\textbf {\bibinfo {volume} {9}},\
  \bibinfo {pages} {1679} (\bibinfo {year} {1997})}\BibitemShut {NoStop}%
\bibitem [{\citenamefont {Catalan}\ \emph
  {et~al.}(2000{\natexlab{a}})\citenamefont {Catalan}, \citenamefont {Bowman},\
  and\ \citenamefont {Gregg}}]{catalan2000transport}%
  \BibitemOpen
  \bibfield  {author} {\bibinfo {author} {\bibfnamefont {G.}~\bibnamefont
  {Catalan}}, \bibinfo {author} {\bibfnamefont {R.}~\bibnamefont {Bowman}}, \
  and\ \bibinfo {author} {\bibfnamefont {J.}~\bibnamefont {Gregg}},\
  }\href@noop {} {\bibfield  {journal} {\bibinfo  {journal} {Journal of Applied
  Physics}\ }\textbf {\bibinfo {volume} {87}},\ \bibinfo {pages} {606}
  (\bibinfo {year} {2000}{\natexlab{a}})}\BibitemShut {NoStop}%
\bibitem [{\citenamefont {Catalan}\ \emph
  {et~al.}(2000{\natexlab{b}})\citenamefont {Catalan}, \citenamefont {Bowman},\
  and\ \citenamefont {Gregg}}]{catalan2000metal}%
  \BibitemOpen
  \bibfield  {author} {\bibinfo {author} {\bibfnamefont {G.}~\bibnamefont
  {Catalan}}, \bibinfo {author} {\bibfnamefont {R.}~\bibnamefont {Bowman}}, \
  and\ \bibinfo {author} {\bibfnamefont {J.}~\bibnamefont {Gregg}},\
  }\href@noop {} {\bibfield  {journal} {\bibinfo  {journal} {Physical Review
  B}\ }\textbf {\bibinfo {volume} {62}},\ \bibinfo {pages} {7892} (\bibinfo
  {year} {2000}{\natexlab{b}})}\BibitemShut {NoStop}%
\bibitem [{\citenamefont {Catalan}(2008)}]{catalan2008progress}%
  \BibitemOpen
  \bibfield  {author} {\bibinfo {author} {\bibfnamefont {G.}~\bibnamefont
  {Catalan}},\ }\href@noop {} {\bibfield  {journal} {\bibinfo  {journal} {Phase
  Transitions}\ }\textbf {\bibinfo {volume} {81}},\ \bibinfo {pages} {729}
  (\bibinfo {year} {2008})}\BibitemShut {NoStop}%
\bibitem [{\citenamefont {Middey}\ \emph {et~al.}(2016)\citenamefont {Middey},
  \citenamefont {Chakhalian}, \citenamefont {Mahadevan}, \citenamefont
  {Freeland}, \citenamefont {Millis},\ and\ \citenamefont
  {Sarma}}]{middey2016physics}%
  \BibitemOpen
  \bibfield  {author} {\bibinfo {author} {\bibfnamefont {S.}~\bibnamefont
  {Middey}}, \bibinfo {author} {\bibfnamefont {J.}~\bibnamefont {Chakhalian}},
  \bibinfo {author} {\bibfnamefont {P.}~\bibnamefont {Mahadevan}}, \bibinfo
  {author} {\bibfnamefont {J.}~\bibnamefont {Freeland}}, \bibinfo {author}
  {\bibfnamefont {A.~J.}\ \bibnamefont {Millis}}, \ and\ \bibinfo {author}
  {\bibfnamefont {D.}~\bibnamefont {Sarma}},\ }\href@noop {} {\bibfield
  {journal} {\bibinfo  {journal} {Annual Review of Materials Research}\
  }\textbf {\bibinfo {volume} {46}},\ \bibinfo {pages} {305} (\bibinfo {year}
  {2016})}\BibitemShut {NoStop}%
\bibitem [{\citenamefont {Catalano}\ \emph {et~al.}(2018)\citenamefont
  {Catalano}, \citenamefont {Gibert}, \citenamefont {Fowlie}, \citenamefont
  {Iniguez}, \citenamefont {Triscone},\ and\ \citenamefont
  {Kreisel}}]{catalano2018rare}%
  \BibitemOpen
  \bibfield  {author} {\bibinfo {author} {\bibfnamefont {S.}~\bibnamefont
  {Catalano}}, \bibinfo {author} {\bibfnamefont {M.}~\bibnamefont {Gibert}},
  \bibinfo {author} {\bibfnamefont {J.}~\bibnamefont {Fowlie}}, \bibinfo
  {author} {\bibfnamefont {J.}~\bibnamefont {Iniguez}}, \bibinfo {author}
  {\bibfnamefont {J.-M.}\ \bibnamefont {Triscone}}, \ and\ \bibinfo {author}
  {\bibfnamefont {J.}~\bibnamefont {Kreisel}},\ }\href@noop {} {\bibfield
  {journal} {\bibinfo  {journal} {Reports on Progress in Physics}\ }\textbf
  {\bibinfo {volume} {81}},\ \bibinfo {pages} {046501} (\bibinfo {year}
  {2018})}\BibitemShut {NoStop}%
\bibitem [{\citenamefont {Jaramillo}\ \emph {et~al.}(2014)\citenamefont
  {Jaramillo}, \citenamefont {Ha}, \citenamefont {Silevitch},\ and\
  \citenamefont {Ramanathan}}]{jaramillo2014origins}%
  \BibitemOpen
  \bibfield  {author} {\bibinfo {author} {\bibfnamefont {R.}~\bibnamefont
  {Jaramillo}}, \bibinfo {author} {\bibfnamefont {S.~D.}\ \bibnamefont {Ha}},
  \bibinfo {author} {\bibfnamefont {D.}~\bibnamefont {Silevitch}}, \ and\
  \bibinfo {author} {\bibfnamefont {S.}~\bibnamefont {Ramanathan}},\
  }\href@noop {} {\bibfield  {journal} {\bibinfo  {journal} {Nature Physics}\
  }\textbf {\bibinfo {volume} {10}},\ \bibinfo {pages} {304} (\bibinfo {year}
  {2014})}\BibitemShut {NoStop}%
\bibitem [{\citenamefont {Zaanen}\ \emph {et~al.}(1985)\citenamefont {Zaanen},
  \citenamefont {Sawatzky},\ and\ \citenamefont {Allen}}]{zaanen1985band}%
  \BibitemOpen
  \bibfield  {author} {\bibinfo {author} {\bibfnamefont {J.}~\bibnamefont
  {Zaanen}}, \bibinfo {author} {\bibfnamefont {G.}~\bibnamefont {Sawatzky}}, \
  and\ \bibinfo {author} {\bibfnamefont {J.}~\bibnamefont {Allen}},\
  }\href@noop {} {\bibfield  {journal} {\bibinfo  {journal} {Physical Review
  Letters}\ }\textbf {\bibinfo {volume} {55}},\ \bibinfo {pages} {418}
  (\bibinfo {year} {1985})}\BibitemShut {NoStop}%
\bibitem [{\citenamefont {Mizokawa}\ \emph {et~al.}(1991)\citenamefont
  {Mizokawa}, \citenamefont {Namatame}, \citenamefont {Fujimori}, \citenamefont
  {Akeyama}, \citenamefont {Kondoh}, \citenamefont {Kuroda},\ and\
  \citenamefont {Kosugi}}]{mizokawa1991origin}%
  \BibitemOpen
  \bibfield  {author} {\bibinfo {author} {\bibfnamefont {T.}~\bibnamefont
  {Mizokawa}}, \bibinfo {author} {\bibfnamefont {H.}~\bibnamefont {Namatame}},
  \bibinfo {author} {\bibfnamefont {A.}~\bibnamefont {Fujimori}}, \bibinfo
  {author} {\bibfnamefont {K.}~\bibnamefont {Akeyama}}, \bibinfo {author}
  {\bibfnamefont {H.}~\bibnamefont {Kondoh}}, \bibinfo {author} {\bibfnamefont
  {H.}~\bibnamefont {Kuroda}}, \ and\ \bibinfo {author} {\bibfnamefont
  {N.}~\bibnamefont {Kosugi}},\ }\href@noop {} {\bibfield  {journal} {\bibinfo
  {journal} {Physical review letters}\ }\textbf {\bibinfo {volume} {67}},\
  \bibinfo {pages} {1638} (\bibinfo {year} {1991})}\BibitemShut {NoStop}%
\bibitem [{\citenamefont {Alonso}\ \emph {et~al.}(2000)\citenamefont {Alonso},
  \citenamefont {Mart{\'\i}nez-Lope}, \citenamefont {Casais}, \citenamefont
  {Garc{\'\i}a-Mu{\~n}oz},\ and\ \citenamefont
  {Fern{\'a}ndez-D{\'\i}az}}]{alonso2000room}%
  \BibitemOpen
  \bibfield  {author} {\bibinfo {author} {\bibfnamefont {J.}~\bibnamefont
  {Alonso}}, \bibinfo {author} {\bibfnamefont {M.}~\bibnamefont
  {Mart{\'\i}nez-Lope}}, \bibinfo {author} {\bibfnamefont {M.}~\bibnamefont
  {Casais}}, \bibinfo {author} {\bibfnamefont {J.}~\bibnamefont
  {Garc{\'\i}a-Mu{\~n}oz}}, \ and\ \bibinfo {author} {\bibfnamefont
  {M.}~\bibnamefont {Fern{\'a}ndez-D{\'\i}az}},\ }\href@noop {} {\bibfield
  {journal} {\bibinfo  {journal} {Physical Review B}\ }\textbf {\bibinfo
  {volume} {61}},\ \bibinfo {pages} {1756} (\bibinfo {year}
  {2000})}\BibitemShut {NoStop}%
\bibitem [{\citenamefont {Khomskii}(2001)}]{khomskii2001unusual}%
  \BibitemOpen
  \bibfield  {author} {\bibinfo {author} {\bibfnamefont {D.}~\bibnamefont
  {Khomskii}},\ }\href@noop {} {\bibfield  {journal} {\bibinfo  {journal}
  {arXiv preprint cond-mat/0101164}\ } (\bibinfo {year} {2001})}\BibitemShut
  {NoStop}%
\bibitem [{\citenamefont {Mazin}\ \emph {et~al.}(2007)\citenamefont {Mazin},
  \citenamefont {Khomskii}, \citenamefont {Lengsdorf}, \citenamefont {Alonso},
  \citenamefont {Marshall}, \citenamefont {Ibberson}, \citenamefont
  {Podlesnyak}, \citenamefont {Mart{\'\i}nez-Lope},\ and\ \citenamefont
  {Abd-Elmeguid}}]{mazin2007charge}%
  \BibitemOpen
  \bibfield  {author} {\bibinfo {author} {\bibfnamefont {I.}~\bibnamefont
  {Mazin}}, \bibinfo {author} {\bibfnamefont {D.}~\bibnamefont {Khomskii}},
  \bibinfo {author} {\bibfnamefont {R.}~\bibnamefont {Lengsdorf}}, \bibinfo
  {author} {\bibfnamefont {J.}~\bibnamefont {Alonso}}, \bibinfo {author}
  {\bibfnamefont {W.}~\bibnamefont {Marshall}}, \bibinfo {author}
  {\bibfnamefont {R.}~\bibnamefont {Ibberson}}, \bibinfo {author}
  {\bibfnamefont {A.}~\bibnamefont {Podlesnyak}}, \bibinfo {author}
  {\bibfnamefont {M.}~\bibnamefont {Mart{\'\i}nez-Lope}}, \ and\ \bibinfo
  {author} {\bibfnamefont {M.}~\bibnamefont {Abd-Elmeguid}},\ }\href@noop {}
  {\bibfield  {journal} {\bibinfo  {journal} {Physical review letters}\
  }\textbf {\bibinfo {volume} {98}},\ \bibinfo {pages} {176406} (\bibinfo
  {year} {2007})}\BibitemShut {NoStop}%
\bibitem [{\citenamefont {Medarde}\ \emph {et~al.}(2009)\citenamefont
  {Medarde}, \citenamefont {Dallera}, \citenamefont {Grioni}, \citenamefont
  {Delley}, \citenamefont {Vernay}, \citenamefont {Mesot}, \citenamefont
  {Sikora}, \citenamefont {Alonso},\ and\ \citenamefont
  {Mart{\'\i}nez-Lope}}]{medarde2009charge}%
  \BibitemOpen
  \bibfield  {author} {\bibinfo {author} {\bibfnamefont {M.}~\bibnamefont
  {Medarde}}, \bibinfo {author} {\bibfnamefont {C.}~\bibnamefont {Dallera}},
  \bibinfo {author} {\bibfnamefont {M.}~\bibnamefont {Grioni}}, \bibinfo
  {author} {\bibfnamefont {B.}~\bibnamefont {Delley}}, \bibinfo {author}
  {\bibfnamefont {F.}~\bibnamefont {Vernay}}, \bibinfo {author} {\bibfnamefont
  {J.}~\bibnamefont {Mesot}}, \bibinfo {author} {\bibfnamefont
  {M.}~\bibnamefont {Sikora}}, \bibinfo {author} {\bibfnamefont
  {J.}~\bibnamefont {Alonso}}, \ and\ \bibinfo {author} {\bibfnamefont
  {M.}~\bibnamefont {Mart{\'\i}nez-Lope}},\ }\href@noop {} {\bibfield
  {journal} {\bibinfo  {journal} {Physical Review B}\ }\textbf {\bibinfo
  {volume} {80}},\ \bibinfo {pages} {245105} (\bibinfo {year}
  {2009})}\BibitemShut {NoStop}%
\bibitem [{\citenamefont {Park}\ \emph {et~al.}(2012)\citenamefont {Park},
  \citenamefont {Millis},\ and\ \citenamefont {Marianetti}}]{park2012site}%
  \BibitemOpen
  \bibfield  {author} {\bibinfo {author} {\bibfnamefont {H.}~\bibnamefont
  {Park}}, \bibinfo {author} {\bibfnamefont {A.~J.}\ \bibnamefont {Millis}}, \
  and\ \bibinfo {author} {\bibfnamefont {C.~A.}\ \bibnamefont {Marianetti}},\
  }\href@noop {} {\bibfield  {journal} {\bibinfo  {journal} {Physical review
  letters}\ }\textbf {\bibinfo {volume} {109}},\ \bibinfo {pages} {156402}
  (\bibinfo {year} {2012})}\BibitemShut {NoStop}%
\bibitem [{\citenamefont {Johnston}\ \emph {et~al.}(2014)\citenamefont
  {Johnston}, \citenamefont {Mukherjee}, \citenamefont {Elfimov}, \citenamefont
  {Berciu},\ and\ \citenamefont {Sawatzky}}]{johnston2014charge}%
  \BibitemOpen
  \bibfield  {author} {\bibinfo {author} {\bibfnamefont {S.}~\bibnamefont
  {Johnston}}, \bibinfo {author} {\bibfnamefont {A.}~\bibnamefont {Mukherjee}},
  \bibinfo {author} {\bibfnamefont {I.}~\bibnamefont {Elfimov}}, \bibinfo
  {author} {\bibfnamefont {M.}~\bibnamefont {Berciu}}, \ and\ \bibinfo {author}
  {\bibfnamefont {G.~A.}\ \bibnamefont {Sawatzky}},\ }\href@noop {} {\bibfield
  {journal} {\bibinfo  {journal} {Physical review letters}\ }\textbf {\bibinfo
  {volume} {112}},\ \bibinfo {pages} {106404} (\bibinfo {year}
  {2014})}\BibitemShut {NoStop}%
\bibitem [{\citenamefont {Varignon}\ \emph {et~al.}(2017)\citenamefont
  {Varignon}, \citenamefont {Grisolia}, \citenamefont {{\'I}{\~n}iguez},
  \citenamefont {Barth{\'e}l{\'e}my},\ and\ \citenamefont
  {Bibes}}]{varignon2017complete}%
  \BibitemOpen
  \bibfield  {author} {\bibinfo {author} {\bibfnamefont {J.}~\bibnamefont
  {Varignon}}, \bibinfo {author} {\bibfnamefont {M.~N.}\ \bibnamefont
  {Grisolia}}, \bibinfo {author} {\bibfnamefont {J.}~\bibnamefont
  {{\'I}{\~n}iguez}}, \bibinfo {author} {\bibfnamefont {A.}~\bibnamefont
  {Barth{\'e}l{\'e}my}}, \ and\ \bibinfo {author} {\bibfnamefont
  {M.}~\bibnamefont {Bibes}},\ }\href@noop {} {\bibfield  {journal} {\bibinfo
  {journal} {npj Quantum Materials}\ }\textbf {\bibinfo {volume} {2}},\
  \bibinfo {pages} {21} (\bibinfo {year} {2017})}\BibitemShut {NoStop}%
\bibitem [{\citenamefont {Bisogni}\ \emph {et~al.}(2016)\citenamefont
  {Bisogni}, \citenamefont {Catalano}, \citenamefont {Green}, \citenamefont
  {Gibert}, \citenamefont {Scherwitzl}, \citenamefont {Huang}, \citenamefont
  {Strocov}, \citenamefont {Zubko}, \citenamefont {Balandeh}, \citenamefont
  {Triscone}, \citenamefont {Sawatzky},\ and\ \citenamefont
  {Schmitt}}]{bisogni2016ground}%
  \BibitemOpen
  \bibfield  {author} {\bibinfo {author} {\bibfnamefont {V.}~\bibnamefont
  {Bisogni}}, \bibinfo {author} {\bibfnamefont {S.}~\bibnamefont {Catalano}},
  \bibinfo {author} {\bibfnamefont {R.~J.}\ \bibnamefont {Green}}, \bibinfo
  {author} {\bibfnamefont {M.}~\bibnamefont {Gibert}}, \bibinfo {author}
  {\bibfnamefont {R.}~\bibnamefont {Scherwitzl}}, \bibinfo {author}
  {\bibfnamefont {Y.}~\bibnamefont {Huang}}, \bibinfo {author} {\bibfnamefont
  {V.~N.}\ \bibnamefont {Strocov}}, \bibinfo {author} {\bibfnamefont
  {P.}~\bibnamefont {Zubko}}, \bibinfo {author} {\bibfnamefont
  {S.}~\bibnamefont {Balandeh}}, \bibinfo {author} {\bibfnamefont {J.-M.}\
  \bibnamefont {Triscone}}, \bibinfo {author} {\bibfnamefont {G.}~\bibnamefont
  {Sawatzky}}, \ and\ \bibinfo {author} {\bibfnamefont {T.}~\bibnamefont
  {Schmitt}},\ }\href@noop {} {\bibfield  {journal} {\bibinfo  {journal}
  {Nature Communications}\ }\textbf {\bibinfo {volume} {7}},\ \bibinfo {pages}
  {13017} (\bibinfo {year} {2016})}\BibitemShut {NoStop}%
\bibitem [{\citenamefont {Green}\ \emph {et~al.}(2016)\citenamefont {Green},
  \citenamefont {Haverkort},\ and\ \citenamefont {Sawatzky}}]{green2016bond}%
  \BibitemOpen
  \bibfield  {author} {\bibinfo {author} {\bibfnamefont {R.~J.}\ \bibnamefont
  {Green}}, \bibinfo {author} {\bibfnamefont {M.~W.}\ \bibnamefont
  {Haverkort}}, \ and\ \bibinfo {author} {\bibfnamefont {G.~A.}\ \bibnamefont
  {Sawatzky}},\ }\href@noop {} {\bibfield  {journal} {\bibinfo  {journal}
  {Physical Review B}\ }\textbf {\bibinfo {volume} {94}},\ \bibinfo {pages}
  {195127} (\bibinfo {year} {2016})}\BibitemShut {NoStop}%
\bibitem [{\citenamefont {Shamblin}\ \emph {et~al.}(2018)\citenamefont
  {Shamblin}, \citenamefont {Heres}, \citenamefont {Zhou}, \citenamefont
  {Sangoro}, \citenamefont {Lang}, \citenamefont {Neuefeind}, \citenamefont
  {Alonso},\ and\ \citenamefont {Johnston}}]{shamblin2018experimental}%
  \BibitemOpen
  \bibfield  {author} {\bibinfo {author} {\bibfnamefont {J.}~\bibnamefont
  {Shamblin}}, \bibinfo {author} {\bibfnamefont {M.}~\bibnamefont {Heres}},
  \bibinfo {author} {\bibfnamefont {H.}~\bibnamefont {Zhou}}, \bibinfo {author}
  {\bibfnamefont {J.}~\bibnamefont {Sangoro}}, \bibinfo {author} {\bibfnamefont
  {M.}~\bibnamefont {Lang}}, \bibinfo {author} {\bibfnamefont {J.}~\bibnamefont
  {Neuefeind}}, \bibinfo {author} {\bibfnamefont {J.}~\bibnamefont {Alonso}}, \
  and\ \bibinfo {author} {\bibfnamefont {S.}~\bibnamefont {Johnston}},\
  }\href@noop {} {\bibfield  {journal} {\bibinfo  {journal} {Nature
  communications}\ }\textbf {\bibinfo {volume} {9}},\ \bibinfo {pages} {86}
  (\bibinfo {year} {2018})}\BibitemShut {NoStop}%
\bibitem [{\citenamefont {Mercy}\ \emph {et~al.}(2017)\citenamefont {Mercy},
  \citenamefont {Bieder}, \citenamefont {{\'I}{\~n}iguez},\ and\ \citenamefont
  {Ghosez}}]{mercy2017structurally}%
  \BibitemOpen
  \bibfield  {author} {\bibinfo {author} {\bibfnamefont {A.}~\bibnamefont
  {Mercy}}, \bibinfo {author} {\bibfnamefont {J.}~\bibnamefont {Bieder}},
  \bibinfo {author} {\bibfnamefont {J.}~\bibnamefont {{\'I}{\~n}iguez}}, \ and\
  \bibinfo {author} {\bibfnamefont {P.}~\bibnamefont {Ghosez}},\ }\href@noop {}
  {\bibfield  {journal} {\bibinfo  {journal} {Nature communications}\ }\textbf
  {\bibinfo {volume} {8}},\ \bibinfo {pages} {1677} (\bibinfo {year}
  {2017})}\BibitemShut {NoStop}%
\bibitem [{\citenamefont {Hansmann}\ \emph {et~al.}(2009)\citenamefont
  {Hansmann}, \citenamefont {Yang}, \citenamefont {Toschi}, \citenamefont
  {Khaliullin}, \citenamefont {Andersen},\ and\ \citenamefont
  {Held}}]{hansmann2009turning}%
  \BibitemOpen
  \bibfield  {author} {\bibinfo {author} {\bibfnamefont {P.}~\bibnamefont
  {Hansmann}}, \bibinfo {author} {\bibfnamefont {X.}~\bibnamefont {Yang}},
  \bibinfo {author} {\bibfnamefont {A.}~\bibnamefont {Toschi}}, \bibinfo
  {author} {\bibfnamefont {G.}~\bibnamefont {Khaliullin}}, \bibinfo {author}
  {\bibfnamefont {O.}~\bibnamefont {Andersen}}, \ and\ \bibinfo {author}
  {\bibfnamefont {K.}~\bibnamefont {Held}},\ }\href@noop {} {\bibfield
  {journal} {\bibinfo  {journal} {Physical Review Letters}\ }\textbf {\bibinfo
  {volume} {103}},\ \bibinfo {pages} {016401} (\bibinfo {year}
  {2009})}\BibitemShut {NoStop}%
\bibitem [{\citenamefont {Li}\ \emph {et~al.}(2019)\citenamefont {Li},
  \citenamefont {Lee}, \citenamefont {Wang}, \citenamefont {Osada},
  \citenamefont {Crossley}, \citenamefont {Lee}, \citenamefont {Cui},
  \citenamefont {Hikita},\ and\ \citenamefont
  {Hwang}}]{danfeng2019superconductivity}%
  \BibitemOpen
  \bibfield  {author} {\bibinfo {author} {\bibfnamefont {D.}~\bibnamefont
  {Li}}, \bibinfo {author} {\bibfnamefont {K.}~\bibnamefont {Lee}}, \bibinfo
  {author} {\bibfnamefont {B.~Y.}\ \bibnamefont {Wang}}, \bibinfo {author}
  {\bibfnamefont {M.}~\bibnamefont {Osada}}, \bibinfo {author} {\bibfnamefont
  {S.}~\bibnamefont {Crossley}}, \bibinfo {author} {\bibfnamefont {H.~R.}\
  \bibnamefont {Lee}}, \bibinfo {author} {\bibfnamefont {Y.}~\bibnamefont
  {Cui}}, \bibinfo {author} {\bibfnamefont {Y.}~\bibnamefont {Hikita}}, \ and\
  \bibinfo {author} {\bibfnamefont {H.~Y.}\ \bibnamefont {Hwang}},\ }\href@noop
  {} {\bibfield  {journal} {\bibinfo  {journal} {Nature}\ }\textbf {\bibinfo
  {volume} {572}},\ \bibinfo {pages} {624} (\bibinfo {year}
  {2019})}\BibitemShut {NoStop}%
\bibitem [{\citenamefont {Blasco}\ \emph {et~al.}(1994)\citenamefont {Blasco},
  \citenamefont {Castro},\ and\ \citenamefont {Garcia}}]{blasco1994structural}%
  \BibitemOpen
  \bibfield  {author} {\bibinfo {author} {\bibfnamefont {J.}~\bibnamefont
  {Blasco}}, \bibinfo {author} {\bibfnamefont {M.}~\bibnamefont {Castro}}, \
  and\ \bibinfo {author} {\bibfnamefont {J.}~\bibnamefont {Garcia}},\
  }\href@noop {} {\bibfield  {journal} {\bibinfo  {journal} {Journal of
  Physics: Condensed Matter}\ }\textbf {\bibinfo {volume} {6}},\ \bibinfo
  {pages} {5875} (\bibinfo {year} {1994})}\BibitemShut {NoStop}%
\bibitem [{\citenamefont {Liu}\ \emph {et~al.}(2013)\citenamefont {Liu},
  \citenamefont {Kargarian}, \citenamefont {Kareev}, \citenamefont {Gray},
  \citenamefont {Ryan}, \citenamefont {Cruz}, \citenamefont {Tahir},
  \citenamefont {Chuang}, \citenamefont {Guo}, \citenamefont {Rondinelli} \emph
  {et~al.}}]{liu2013heterointerface}%
  \BibitemOpen
  \bibfield  {author} {\bibinfo {author} {\bibfnamefont {J.}~\bibnamefont
  {Liu}}, \bibinfo {author} {\bibfnamefont {M.}~\bibnamefont {Kargarian}},
  \bibinfo {author} {\bibfnamefont {M.}~\bibnamefont {Kareev}}, \bibinfo
  {author} {\bibfnamefont {B.}~\bibnamefont {Gray}}, \bibinfo {author}
  {\bibfnamefont {P.~J.}\ \bibnamefont {Ryan}}, \bibinfo {author}
  {\bibfnamefont {A.}~\bibnamefont {Cruz}}, \bibinfo {author} {\bibfnamefont
  {N.}~\bibnamefont {Tahir}}, \bibinfo {author} {\bibfnamefont {Y.-D.}\
  \bibnamefont {Chuang}}, \bibinfo {author} {\bibfnamefont {J.}~\bibnamefont
  {Guo}}, \bibinfo {author} {\bibfnamefont {J.~M.}\ \bibnamefont {Rondinelli}},
   \emph {et~al.},\ }\href@noop {} {\bibfield  {journal} {\bibinfo  {journal}
  {Nature communications}\ }\textbf {\bibinfo {volume} {4}},\ \bibinfo {pages}
  {2714} (\bibinfo {year} {2013})}\BibitemShut {NoStop}%
\bibitem [{\citenamefont {Mikheev}\ \emph {et~al.}(2015)\citenamefont
  {Mikheev}, \citenamefont {Hauser}, \citenamefont {Himmetoglu}, \citenamefont
  {Moreno}, \citenamefont {Janotti}, \citenamefont {Van~de Walle},\ and\
  \citenamefont {Stemmer}}]{mikheev2015tuning}%
  \BibitemOpen
  \bibfield  {author} {\bibinfo {author} {\bibfnamefont {E.}~\bibnamefont
  {Mikheev}}, \bibinfo {author} {\bibfnamefont {A.~J.}\ \bibnamefont {Hauser}},
  \bibinfo {author} {\bibfnamefont {B.}~\bibnamefont {Himmetoglu}}, \bibinfo
  {author} {\bibfnamefont {N.~E.}\ \bibnamefont {Moreno}}, \bibinfo {author}
  {\bibfnamefont {A.}~\bibnamefont {Janotti}}, \bibinfo {author} {\bibfnamefont
  {C.~G.}\ \bibnamefont {Van~de Walle}}, \ and\ \bibinfo {author}
  {\bibfnamefont {S.}~\bibnamefont {Stemmer}},\ }\href@noop {} {\bibfield
  {journal} {\bibinfo  {journal} {Science advances}\ }\textbf {\bibinfo
  {volume} {1}},\ \bibinfo {pages} {e1500797} (\bibinfo {year}
  {2015})}\BibitemShut {NoStop}%
\bibitem [{\citenamefont {Kobayashi}\ \emph {et~al.}(2015)\citenamefont
  {Kobayashi}, \citenamefont {Ikeda}, \citenamefont {Yoda}, \citenamefont
  {Hirao}, \citenamefont {Ohishi}, \citenamefont {Alonso}, \citenamefont
  {Martinez-Lope}, \citenamefont {Lengsdorf}, \citenamefont {Khomskii},\ and\
  \citenamefont {Abd-Elmeguid}}]{kobayashi2015pressure}%
  \BibitemOpen
  \bibfield  {author} {\bibinfo {author} {\bibfnamefont {H.}~\bibnamefont
  {Kobayashi}}, \bibinfo {author} {\bibfnamefont {S.}~\bibnamefont {Ikeda}},
  \bibinfo {author} {\bibfnamefont {Y.}~\bibnamefont {Yoda}}, \bibinfo {author}
  {\bibfnamefont {N.}~\bibnamefont {Hirao}}, \bibinfo {author} {\bibfnamefont
  {Y.}~\bibnamefont {Ohishi}}, \bibinfo {author} {\bibfnamefont
  {J.}~\bibnamefont {Alonso}}, \bibinfo {author} {\bibfnamefont
  {M.}~\bibnamefont {Martinez-Lope}}, \bibinfo {author} {\bibfnamefont
  {R.}~\bibnamefont {Lengsdorf}}, \bibinfo {author} {\bibfnamefont
  {D.}~\bibnamefont {Khomskii}}, \ and\ \bibinfo {author} {\bibfnamefont
  {M.}~\bibnamefont {Abd-Elmeguid}},\ }\href@noop {} {\bibfield  {journal}
  {\bibinfo  {journal} {Physical Review B}\ }\textbf {\bibinfo {volume} {91}},\
  \bibinfo {pages} {195148} (\bibinfo {year} {2015})}\BibitemShut {NoStop}%
\bibitem [{\citenamefont {Yadav}\ \emph {et~al.}(2018)\citenamefont {Yadav},
  \citenamefont {Harisankar}, \citenamefont {Soni},\ and\ \citenamefont
  {Mavani}}]{yadav2018influence}%
  \BibitemOpen
  \bibfield  {author} {\bibinfo {author} {\bibfnamefont {E.}~\bibnamefont
  {Yadav}}, \bibinfo {author} {\bibfnamefont {S.}~\bibnamefont {Harisankar}},
  \bibinfo {author} {\bibfnamefont {K.}~\bibnamefont {Soni}}, \ and\ \bibinfo
  {author} {\bibfnamefont {K.}~\bibnamefont {Mavani}},\ }\href@noop {}
  {\bibfield  {journal} {\bibinfo  {journal} {Applied Physics A}\ }\textbf
  {\bibinfo {volume} {124}},\ \bibinfo {pages} {614} (\bibinfo {year}
  {2018})}\BibitemShut {NoStop}%
\bibitem [{\citenamefont {Phanindra}\ \emph {et~al.}(2018)\citenamefont
  {Phanindra}, \citenamefont {Agarwal},\ and\ \citenamefont
  {Rana}}]{phanindra2018terahertz}%
  \BibitemOpen
  \bibfield  {author} {\bibinfo {author} {\bibfnamefont {V.~E.}\ \bibnamefont
  {Phanindra}}, \bibinfo {author} {\bibfnamefont {P.}~\bibnamefont {Agarwal}},
  \ and\ \bibinfo {author} {\bibfnamefont {D.}~\bibnamefont {Rana}},\
  }\href@noop {} {\bibfield  {journal} {\bibinfo  {journal} {Physical Review
  Materials}\ }\textbf {\bibinfo {volume} {2}},\ \bibinfo {pages} {015001}
  (\bibinfo {year} {2018})}\BibitemShut {NoStop}%
\bibitem [{\citenamefont {Hussey}\ \emph {et~al.}(2004)\citenamefont {Hussey},
  \citenamefont {Takenaka},\ and\ \citenamefont
  {Takagi}}]{hussey2004universality}%
  \BibitemOpen
  \bibfield  {author} {\bibinfo {author} {\bibfnamefont {N.}~\bibnamefont
  {Hussey}}, \bibinfo {author} {\bibfnamefont {K.}~\bibnamefont {Takenaka}}, \
  and\ \bibinfo {author} {\bibfnamefont {H.}~\bibnamefont {Takagi}},\
  }\href@noop {} {\bibfield  {journal} {\bibinfo  {journal} {Philosophical
  Magazine}\ }\textbf {\bibinfo {volume} {84}},\ \bibinfo {pages} {2847}
  (\bibinfo {year} {2004})}\BibitemShut {NoStop}%
\bibitem [{\citenamefont {Patel}\ \emph {et~al.}(2017)\citenamefont {Patel},
  \citenamefont {Mukherjee}, \citenamefont {Kaushal}, \citenamefont {Moreo},\
  and\ \citenamefont {Dagotto}}]{patel2017non}%
  \BibitemOpen
  \bibfield  {author} {\bibinfo {author} {\bibfnamefont {N.~D.}\ \bibnamefont
  {Patel}}, \bibinfo {author} {\bibfnamefont {A.}~\bibnamefont {Mukherjee}},
  \bibinfo {author} {\bibfnamefont {N.}~\bibnamefont {Kaushal}}, \bibinfo
  {author} {\bibfnamefont {A.}~\bibnamefont {Moreo}}, \ and\ \bibinfo {author}
  {\bibfnamefont {E.}~\bibnamefont {Dagotto}},\ }\href@noop {} {\bibfield
  {journal} {\bibinfo  {journal} {Physical Review Letters}\ }\textbf {\bibinfo
  {volume} {119}},\ \bibinfo {pages} {086601} (\bibinfo {year}
  {2017})}\BibitemShut {NoStop}%
\bibitem [{\citenamefont {Preziosi}\ \emph {et~al.}(2017)\citenamefont
  {Preziosi}, \citenamefont {Sander}, \citenamefont {Barth{\'e}l{\'e}my},\ and\
  \citenamefont {Bibes}}]{preziosi2017reproducibility}%
  \BibitemOpen
  \bibfield  {author} {\bibinfo {author} {\bibfnamefont {D.}~\bibnamefont
  {Preziosi}}, \bibinfo {author} {\bibfnamefont {A.}~\bibnamefont {Sander}},
  \bibinfo {author} {\bibfnamefont {A.}~\bibnamefont {Barth{\'e}l{\'e}my}}, \
  and\ \bibinfo {author} {\bibfnamefont {M.}~\bibnamefont {Bibes}},\
  }\href@noop {} {\bibfield  {journal} {\bibinfo  {journal} {AIP Advances}\
  }\textbf {\bibinfo {volume} {7}},\ \bibinfo {pages} {015210} (\bibinfo {year}
  {2017})}\BibitemShut {NoStop}%
\bibitem [{\citenamefont {Salamon}\ \emph {et~al.}(2002)\citenamefont
  {Salamon}, \citenamefont {Lin},\ and\ \citenamefont
  {Chun}}]{Salamon2002Colossal}%
  \BibitemOpen
  \bibfield  {author} {\bibinfo {author} {\bibfnamefont {M.}~\bibnamefont
  {Salamon}}, \bibinfo {author} {\bibfnamefont {P.}~\bibnamefont {Lin}}, \ and\
  \bibinfo {author} {\bibfnamefont {S.~H.}\ \bibnamefont {Chun}},\ }\href@noop
  {} {\bibfield  {journal} {\bibinfo  {journal} {Physical Review Letters}\
  }\textbf {\bibinfo {volume} {88}},\ \bibinfo {pages} {1972013} (\bibinfo
  {year} {2002})}\BibitemShut {NoStop}%
\bibitem [{\citenamefont {Bruin}\ \emph {et~al.}(2013)\citenamefont {Bruin},
  \citenamefont {Sakai}, \citenamefont {Perry},\ and\ \citenamefont
  {Mackenzie}}]{bruin2013similarity}%
  \BibitemOpen
  \bibfield  {author} {\bibinfo {author} {\bibfnamefont {J.}~\bibnamefont
  {Bruin}}, \bibinfo {author} {\bibfnamefont {H.}~\bibnamefont {Sakai}},
  \bibinfo {author} {\bibfnamefont {R.}~\bibnamefont {Perry}}, \ and\ \bibinfo
  {author} {\bibfnamefont {A.}~\bibnamefont {Mackenzie}},\ }\href@noop {}
  {\bibfield  {journal} {\bibinfo  {journal} {Science}\ }\textbf {\bibinfo
  {volume} {339}},\ \bibinfo {pages} {804} (\bibinfo {year}
  {2013})}\BibitemShut {NoStop}%
\bibitem [{\citenamefont {Bak}\ \emph {et~al.}(2017)\citenamefont {Bak},
  \citenamefont {Bae}, \citenamefont {Kim}, \citenamefont {Oh},\ and\
  \citenamefont {Chung}}]{bak2017formation}%
  \BibitemOpen
  \bibfield  {author} {\bibinfo {author} {\bibfnamefont {J.}~\bibnamefont
  {Bak}}, \bibinfo {author} {\bibfnamefont {H.~B.}\ \bibnamefont {Bae}},
  \bibinfo {author} {\bibfnamefont {J.}~\bibnamefont {Kim}}, \bibinfo {author}
  {\bibfnamefont {J.}~\bibnamefont {Oh}}, \ and\ \bibinfo {author}
  {\bibfnamefont {S.-Y.}\ \bibnamefont {Chung}},\ }\href@noop {} {\bibfield
  {journal} {\bibinfo  {journal} {Nano letters}\ }\textbf {\bibinfo {volume}
  {17}},\ \bibinfo {pages} {3126} (\bibinfo {year} {2017})}\BibitemShut
  {NoStop}%
\bibitem [{\citenamefont {Coll}\ \emph {et~al.}(2017)\citenamefont {Coll},
  \citenamefont {L{\'o}pez-Conesa}, \citenamefont {Rebled}, \citenamefont
  {Mag{\'e}n}, \citenamefont {Sanchez}, \citenamefont {Fontcuberta},
  \citenamefont {Estrad},\ and\ \citenamefont
  {Francesca}}]{coll2017simulation}%
  \BibitemOpen
  \bibfield  {author} {\bibinfo {author} {\bibfnamefont {C.}~\bibnamefont
  {Coll}}, \bibinfo {author} {\bibfnamefont {L.}~\bibnamefont
  {L{\'o}pez-Conesa}}, \bibinfo {author} {\bibfnamefont {J.~M.}\ \bibnamefont
  {Rebled}}, \bibinfo {author} {\bibfnamefont {C.}~\bibnamefont {Mag{\'e}n}},
  \bibinfo {author} {\bibfnamefont {F.}~\bibnamefont {Sanchez}}, \bibinfo
  {author} {\bibfnamefont {J.}~\bibnamefont {Fontcuberta}}, \bibinfo {author}
  {\bibfnamefont {S.}~\bibnamefont {Estrad}}, \ and\ \bibinfo {author}
  {\bibnamefont {Francesca}},\ }\href@noop {} {\bibfield  {journal} {\bibinfo
  {journal} {The Journal of Physical Chemistry C}\ }\textbf {\bibinfo {volume}
  {121}},\ \bibinfo {pages} {9300} (\bibinfo {year} {2017})}\BibitemShut
  {NoStop}%
\bibitem [{\citenamefont {L{\'o}pez-Conesa}\ \emph {et~al.}(2017)\citenamefont
  {L{\'o}pez-Conesa}, \citenamefont {Rebled}, \citenamefont {Pesquera},
  \citenamefont {Dix}, \citenamefont {S{\'a}nchez}, \citenamefont {Herranz},
  \citenamefont {Fontcuberta}, \citenamefont {Mag{\'e}n}, \citenamefont
  {Casanove}, \citenamefont {Estrad{\'e}} \emph {et~al.}}]{lopez2017evidence}%
  \BibitemOpen
  \bibfield  {author} {\bibinfo {author} {\bibfnamefont {L.}~\bibnamefont
  {L{\'o}pez-Conesa}}, \bibinfo {author} {\bibfnamefont {J.~M.}\ \bibnamefont
  {Rebled}}, \bibinfo {author} {\bibfnamefont {D.}~\bibnamefont {Pesquera}},
  \bibinfo {author} {\bibfnamefont {N.}~\bibnamefont {Dix}}, \bibinfo {author}
  {\bibfnamefont {F.}~\bibnamefont {S{\'a}nchez}}, \bibinfo {author}
  {\bibfnamefont {G.}~\bibnamefont {Herranz}}, \bibinfo {author} {\bibfnamefont
  {J.}~\bibnamefont {Fontcuberta}}, \bibinfo {author} {\bibfnamefont
  {C.}~\bibnamefont {Mag{\'e}n}}, \bibinfo {author} {\bibfnamefont {M.~J.}\
  \bibnamefont {Casanove}}, \bibinfo {author} {\bibfnamefont {S.}~\bibnamefont
  {Estrad{\'e}}},  \emph {et~al.},\ }\href@noop {} {\bibfield  {journal}
  {\bibinfo  {journal} {Physical Chemistry Chemical Physics}\ }\textbf
  {\bibinfo {volume} {19}},\ \bibinfo {pages} {9137} (\bibinfo {year}
  {2017})}\BibitemShut {NoStop}%
\bibitem [{\citenamefont {Malashevich}\ and\ \citenamefont
  {Ismail-Beigi}(2015)}]{malashevich2015first}%
  \BibitemOpen
  \bibfield  {author} {\bibinfo {author} {\bibfnamefont {A.}~\bibnamefont
  {Malashevich}}\ and\ \bibinfo {author} {\bibfnamefont {S.}~\bibnamefont
  {Ismail-Beigi}},\ }\href@noop {} {\bibfield  {journal} {\bibinfo  {journal}
  {Physical Review B}\ }\textbf {\bibinfo {volume} {92}},\ \bibinfo {pages}
  {144102} (\bibinfo {year} {2015})}\BibitemShut {NoStop}%
\bibitem [{\citenamefont {Wang}\ \emph {et~al.}(2016)\citenamefont {Wang},
  \citenamefont {Dash}, \citenamefont {Chang}, \citenamefont {You},
  \citenamefont {Feng}, \citenamefont {He}, \citenamefont {Jin}, \citenamefont
  {Zhou}, \citenamefont {Ong}, \citenamefont {Ren} \emph
  {et~al.}}]{wang2016oxygen}%
  \BibitemOpen
  \bibfield  {author} {\bibinfo {author} {\bibfnamefont {L.}~\bibnamefont
  {Wang}}, \bibinfo {author} {\bibfnamefont {S.}~\bibnamefont {Dash}}, \bibinfo
  {author} {\bibfnamefont {L.}~\bibnamefont {Chang}}, \bibinfo {author}
  {\bibfnamefont {L.}~\bibnamefont {You}}, \bibinfo {author} {\bibfnamefont
  {Y.}~\bibnamefont {Feng}}, \bibinfo {author} {\bibfnamefont {X.}~\bibnamefont
  {He}}, \bibinfo {author} {\bibfnamefont {K.-j.}\ \bibnamefont {Jin}},
  \bibinfo {author} {\bibfnamefont {Y.}~\bibnamefont {Zhou}}, \bibinfo {author}
  {\bibfnamefont {H.~G.}\ \bibnamefont {Ong}}, \bibinfo {author} {\bibfnamefont
  {P.}~\bibnamefont {Ren}},  \emph {et~al.},\ }\href@noop {} {\bibfield
  {journal} {\bibinfo  {journal} {ACS applied materials \& interfaces}\
  }\textbf {\bibinfo {volume} {8}},\ \bibinfo {pages} {9769} (\bibinfo {year}
  {2016})}\BibitemShut {NoStop}%
\bibitem [{\citenamefont {Hauser}\ \emph {et~al.}(2015)\citenamefont {Hauser},
  \citenamefont {Mikheev}, \citenamefont {Moreno}, \citenamefont {Hwang},
  \citenamefont {Zhang},\ and\ \citenamefont
  {Stemmer}}]{hauser2015correlation}%
  \BibitemOpen
  \bibfield  {author} {\bibinfo {author} {\bibfnamefont {A.~J.}\ \bibnamefont
  {Hauser}}, \bibinfo {author} {\bibfnamefont {E.}~\bibnamefont {Mikheev}},
  \bibinfo {author} {\bibfnamefont {N.~E.}\ \bibnamefont {Moreno}}, \bibinfo
  {author} {\bibfnamefont {J.}~\bibnamefont {Hwang}}, \bibinfo {author}
  {\bibfnamefont {J.~Y.}\ \bibnamefont {Zhang}}, \ and\ \bibinfo {author}
  {\bibfnamefont {S.}~\bibnamefont {Stemmer}},\ }\href@noop {} {\bibfield
  {journal} {\bibinfo  {journal} {Applied Physics Letters}\ }\textbf {\bibinfo
  {volume} {106}},\ \bibinfo {pages} {092104} (\bibinfo {year}
  {2015})}\BibitemShut {NoStop}%
\bibitem [{\citenamefont {Iglesias}\ \emph {et~al.}(2017)\citenamefont
  {Iglesias}, \citenamefont {Sarantopoulos}, \citenamefont {Mag{\'e}n},\ and\
  \citenamefont {Rivadulla}}]{iglesias2017oxygen}%
  \BibitemOpen
  \bibfield  {author} {\bibinfo {author} {\bibfnamefont {L.}~\bibnamefont
  {Iglesias}}, \bibinfo {author} {\bibfnamefont {A.}~\bibnamefont
  {Sarantopoulos}}, \bibinfo {author} {\bibfnamefont {C.}~\bibnamefont
  {Mag{\'e}n}}, \ and\ \bibinfo {author} {\bibfnamefont {F.}~\bibnamefont
  {Rivadulla}},\ }\href@noop {} {\bibfield  {journal} {\bibinfo  {journal}
  {Physical Review B}\ }\textbf {\bibinfo {volume} {95}},\ \bibinfo {pages}
  {165138} (\bibinfo {year} {2017})}\BibitemShut {NoStop}%
\bibitem [{\citenamefont {Herranz}\ \emph {et~al.}(2008)\citenamefont
  {Herranz}, \citenamefont {Laukhin}, \citenamefont {S{\'a}nchez},
  \citenamefont {Levy}, \citenamefont {Ferrater}, \citenamefont
  {Garc{\'\i}a-Cuenca}, \citenamefont {Varela},\ and\ \citenamefont
  {Fontcuberta}}]{herranz2008effect}%
  \BibitemOpen
  \bibfield  {author} {\bibinfo {author} {\bibfnamefont {G.}~\bibnamefont
  {Herranz}}, \bibinfo {author} {\bibfnamefont {V.}~\bibnamefont {Laukhin}},
  \bibinfo {author} {\bibfnamefont {F.}~\bibnamefont {S{\'a}nchez}}, \bibinfo
  {author} {\bibfnamefont {P.}~\bibnamefont {Levy}}, \bibinfo {author}
  {\bibfnamefont {C.}~\bibnamefont {Ferrater}}, \bibinfo {author}
  {\bibfnamefont {M.}~\bibnamefont {Garc{\'\i}a-Cuenca}}, \bibinfo {author}
  {\bibfnamefont {M.}~\bibnamefont {Varela}}, \ and\ \bibinfo {author}
  {\bibfnamefont {J.}~\bibnamefont {Fontcuberta}},\ }\href@noop {} {\bibfield
  {journal} {\bibinfo  {journal} {Physical Review B}\ }\textbf {\bibinfo
  {volume} {77}},\ \bibinfo {pages} {165114} (\bibinfo {year}
  {2008})}\BibitemShut {NoStop}%
\bibitem [{\citenamefont {Mott}(1969)}]{mott1969conduction}%
  \BibitemOpen
  \bibfield  {author} {\bibinfo {author} {\bibfnamefont {N.~F.}\ \bibnamefont
  {Mott}},\ }\href@noop {} {\bibfield  {journal} {\bibinfo  {journal}
  {Philosophical Magazine}\ }\textbf {\bibinfo {volume} {19}},\ \bibinfo
  {pages} {835} (\bibinfo {year} {1969})}\BibitemShut {NoStop}%
\bibitem [{\citenamefont {Issai}(2015)}]{issai2015hopping}%
  \BibitemOpen
  \bibfield  {author} {\bibinfo {author} {\bibfnamefont {S.}~\bibnamefont
  {Issai}},\ }\href@noop {} {\emph {\bibinfo {title} {Is hopping a science?:
  selected topics of hopping conductivity}}}\ (\bibinfo  {publisher} {World
  Scientific},\ \bibinfo {year} {2015})\BibitemShut {NoStop}%
\bibitem [{\citenamefont {Wang}\ \emph {et~al.}(2018)\citenamefont {Wang},
  \citenamefont {Rosenkranz}, \citenamefont {Rui}, \citenamefont {Zhang},
  \citenamefont {Ye}, \citenamefont {Zheng}, \citenamefont {Klie},
  \citenamefont {Mitchell},\ and\ \citenamefont
  {Phelan}}]{wang2018antiferromagnetic}%
  \BibitemOpen
  \bibfield  {author} {\bibinfo {author} {\bibfnamefont {B.-X.}\ \bibnamefont
  {Wang}}, \bibinfo {author} {\bibfnamefont {S.}~\bibnamefont {Rosenkranz}},
  \bibinfo {author} {\bibfnamefont {X.}~\bibnamefont {Rui}}, \bibinfo {author}
  {\bibfnamefont {J.}~\bibnamefont {Zhang}}, \bibinfo {author} {\bibfnamefont
  {F.}~\bibnamefont {Ye}}, \bibinfo {author} {\bibfnamefont {H.}~\bibnamefont
  {Zheng}}, \bibinfo {author} {\bibfnamefont {R.}~\bibnamefont {Klie}},
  \bibinfo {author} {\bibfnamefont {J.~F.}\ \bibnamefont {Mitchell}}, \ and\
  \bibinfo {author} {\bibfnamefont {D.}~\bibnamefont {Phelan}},\ }\href@noop {}
  {\bibfield  {journal} {\bibinfo  {journal} {Physical Review Materials}\
  }\textbf {\bibinfo {volume} {2}},\ \bibinfo {pages} {064404} (\bibinfo {year}
  {2018})}\BibitemShut {NoStop}%
\end{thebibliography}%

\clearpage

\section{Supplementary Material}

\subsection{Thin film growth}

Epitaxial NdNiO\textsubscript{3} thin films were deposited on single-crystal  LaAlO\textsubscript{3} (LAO),  NdGaO\textsubscript{3} (NGO),  SrTiO\textsubscript{3} (STO) and  DyScO\textsubscript{3} (DSO) substrates by pulsed laser ablation of a single-phase target (Toshima Manufacturing Co., Ltd.). The quality of the target is of crucial importance to attain reproducibility of the film properties. Before deposition, the LAO substrates were thermally annealed at 1050 $^\circ$C in a flow of O\textsubscript{2} and etched with DI water to obtain an atomically flat surface with single terminated terraces. The NGO and STO substrates were etched with buffered NH\textsubscript{4}F (10 M)-HF solution (BHF) and the DSO substrates were etched with NaOH. All the substrates displayed single terminated terraces after the treatment. The substrates were heated to a temperature of 700 $^\circ$C prior to the deposition of the films and were kept at that temperature during growth. Oxygen was present in the growth chamber during deposition with an oxygen pressure of 0.2 mbar and the laser fluence on the target was 2 J/cm\textsuperscript{2}. After deposition, the samples were cooled down to room temperature at 5 $^\circ$C/min. The growth was monitored using Reflection High Energy Electron Diffraction (RHEED). The films showed a constant deposition time of about 22 s per unit cell (s/uc) for NNO/LAO and 24 s/uc for  NNO/STO. Films with various thicknesses were grown by precisely tuning the deposition time.

\subsection{Characterization methods}

The structural and transport properties of all films were studied in detail. The in-situ RHEED patterns and atomic force microscopy (AFM) topography images were captured for the as-grown films, as shown in Fig. \ref{fig:RHEED-AFM} for one of the films. The intensity oscillations of NNO/LAO and NNO/STO films indicate that at least the first 13 layers ($\sim$ 5 nm) of NNO film are deposited atomic-layer by atomic-layer. 

Cross-sectional specimens of the films were prepared and studied by scanning transmission electron microscopy (STEM) on a probe corrected FEI Titan 60–300 microscope equipped with a high-brightness field emission gun (X-FEG) and a CEOS aberration corrector for the condenser system. This microscope was operated at 300 kV. High angle annular dark field  (HAADF) STEM images were acquired with a convergence angle of 25 mrad and a probe size below 1 Å. The strain state of the films was determined by geometrical phase analysis (GPA) of these HAADF images.

Transport properties were measured between 5 K to 400 K by the van der Pauw method in a Quantum Design Physical Property Measurement System (PPMS), using a Keithley 237 current source and a Agilent 3458A multimeter.

\subsection{Extracting the exponent \textit{n}}

It is important to pay attention to the extraction of the $\rho$(T) power law exponents from the experimental data. Proper determination of scaling exponents requires access to many scales in the experimental parameters. In the case of temperature scaling, this is not possible so the exponents obtained from the experimental resistivity-temperature analysis can be considered as \textit{apparent} scaling exponents. In order to extract these exponents, the experimentally obtained $\rho$(T) of each film was plotted as a function of \textit{T}\textsuperscript{n} for different \textit{n} values, as shown for one of the films in Fig. \ref{fig:n}a. All the different curves were fitted as linear fits and the best fit (the best \textit{n}) was determined by the largest R$^2$ factor. This was done by interpolating to the maximum of the R$^2$ parabola, as shown for several examples in Fig. \ref{fig:n}b-f.

The range of validity of the fits was checked by plotting the temperature derivative of $ \rho$(T) as a function \textit{T}$^{n}$. These plots can be seen for some of the films in Fig. \ref{fig:resis2}.

\subsection{NdNiO$_3$ thin films on NdGaO$_3$ substrate}

Next to the series of samples deposited on LAO and STO substrates, NNO films with 5 nm in thickness have been grown on NdGaO$_3$ substrates, imposing a nominal tensile strain value of + 1.35$\%$. These films show transport properties similar to those of the films on STO under similar tensile strain (see Fig. S4a).

\subsection{NdNiO$_3$ thin films on DyScO$_3$ substrate}

As mentioned in the main text, large enough tensile strain should induce a large density of defects that would, eventually, suppress the metallic phase. This is, indeed, shown in the inset of Fig. S5a for film grown on DSO substrates (+3.86 $\%$ tensile strain). There is no metallic phase below 400 K. In this case, the resistivity is well described by a Mott's variable range hopping (VRH) conduction model (see Fig. S5b) for T$<$ 80 K; while the data follows a simple thermal activation model with a single activation energy E$_a$= 32 meV (consistent with a near neighbours hopping conductions model, NNH) for temperatures above T= 80 K. This evolution from VRH to NNH with increasing temperature is characteristic of disordered solids.

\subsection{Intermediate state of NNO/STO film}

In Fig. S6, the resistivity data of a 10 nm film on DSO (same data as in Fig. S5) is plotted together with the fit to the NNH model with E$_a$= 32 meV, which is valid for temperature above about 70-80 K. In addition, the data for another two films of the same thickness under low strain (on LAO) and intermediate strain (on STO) are also shown. The film on STO displays similar behaviour as the film on DSO in the insulating phase: a NNH regime at T$>$ 80 K. It is interesting to notice that these films under intermediate strain (on STO) show a magnitude of the resistivity in the metallic state and a T$_{MI}$ that are in between those of the film on DSO and the film on LAO. In addition, the slope of the resistivity of the film on STO seems to approach that of the film on LAO at the largest temperatures.

\renewcommand{\thefigure}{S1}
\begin{figure*}[h!]
\centering
\includegraphics[scale=0.2]{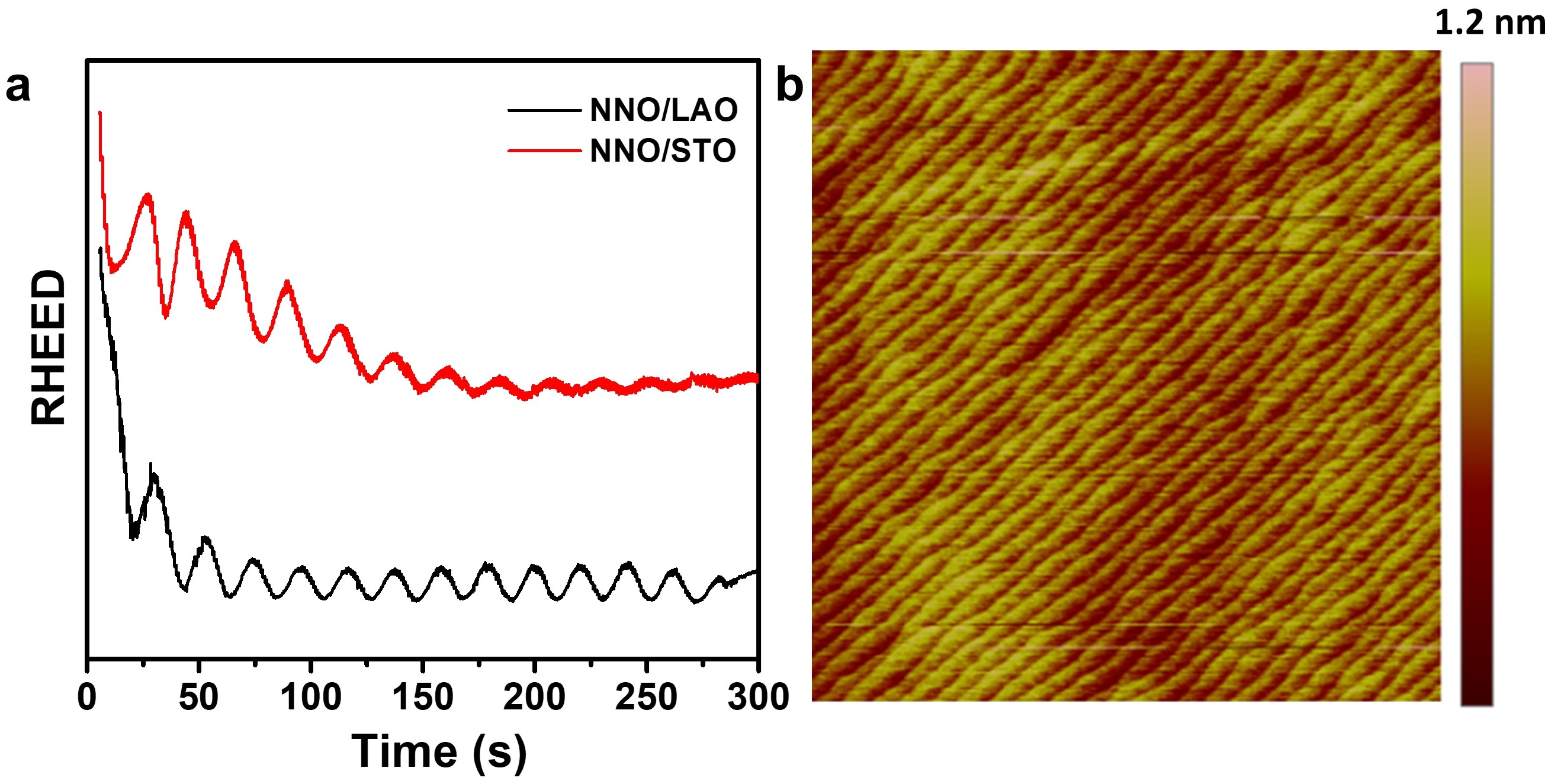}
\caption{\label{fig:RHEED-AFM}(a) RHEED intensity oscillations observed \textit{in-situ} during PLD deposition. (b) AFM topography image of a 5$\times$5 $\mu$ m$\times$ $\mu$ m area in a 5 nm thick NNO film grown on a LAO substrate.}
\end{figure*}

\renewcommand{\thefigure}{S2}
\begin{figure*}[h!]
\centering
\includegraphics[scale=0.2]{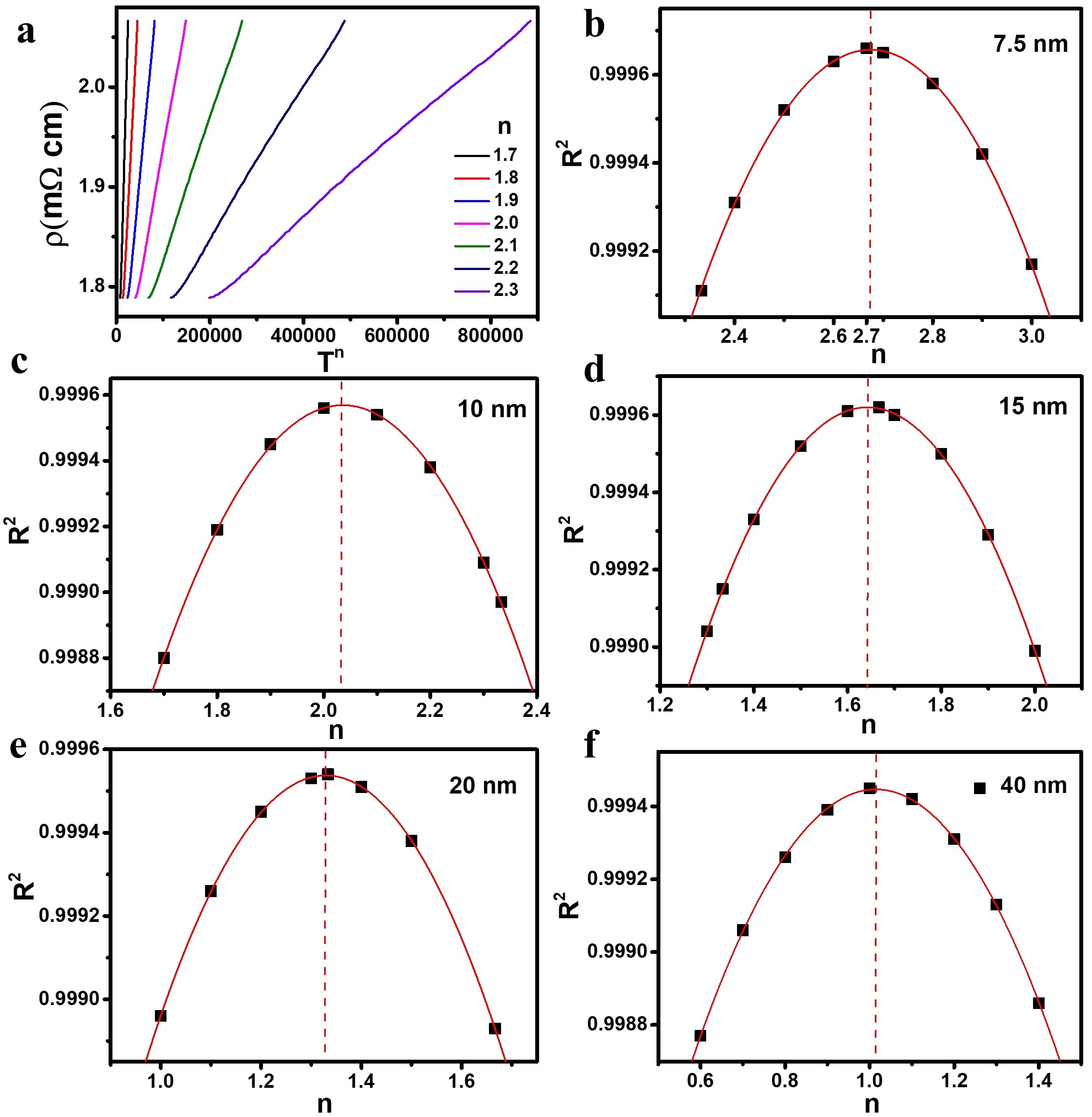}
\caption{\label{fig:n} Resistivity of a 10 nm NNO/STO film as a function of T\textsuperscript{n} with different n values.  (b-f) The coefficient of determination (R2) as a function ofnfor NNO/STO films with thicknesses of 7.5 nm, 10 nm, 15 nm, 20 nm and 40nm, respectively.}
\end{figure*}

\renewcommand{\thefigure}{S3}
\begin{figure*}[h!]
\centering
\includegraphics[scale=0.2]{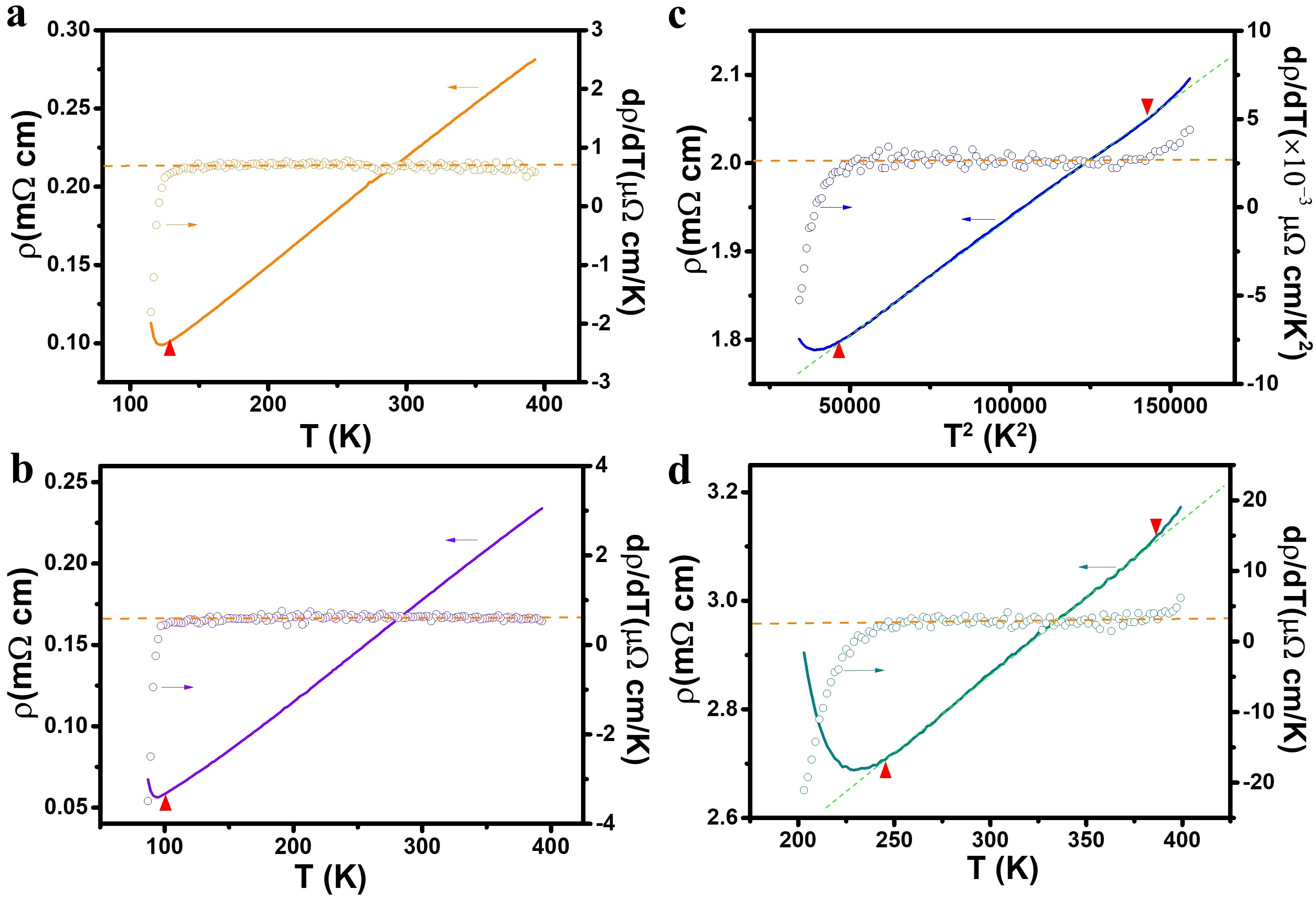}
\caption{\label{fig:resis2} Resistivity (solid lines) and its temperature derivative (open circles) versus T$^n$ for NNO/LAO films with thickness of 5 nm (a) and 7.5 nm (b) and for NNO/STO films with thicknesses of 10 nm (c) and 40 nm (d). The upturn in resistivity signalling the metal-insulator transition is visible. The red triangles signal the temperature region where the resistivity starts to deviate from the best fit.}
\end{figure*}

\renewcommand{\thefigure}{S4}
\begin{figure*}[h!]
\centering
\includegraphics[scale=0.22]{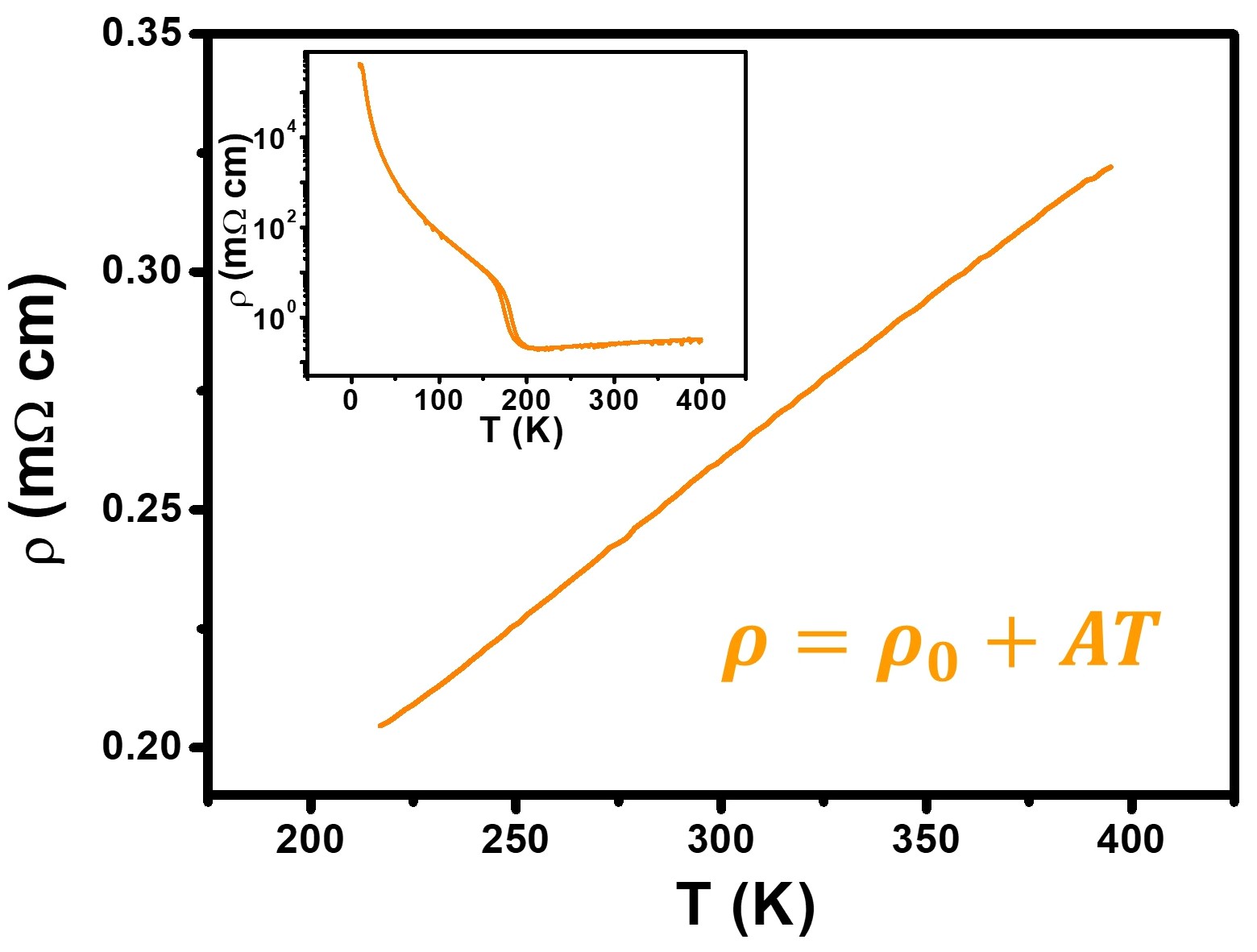}
\caption{\label{fig:onNGO}  \textit{T}-linear dependent resistivity in the metallic phase of a 5 nm NNO/NGO film. Inset: Temperature dependence of resistivity in an extended temperature range, including both cooling and heating processes.}
\end{figure*}

\renewcommand{\thefigure}{S5}
\begin{figure*}[h!]
\centering
\includegraphics[scale=0.22]{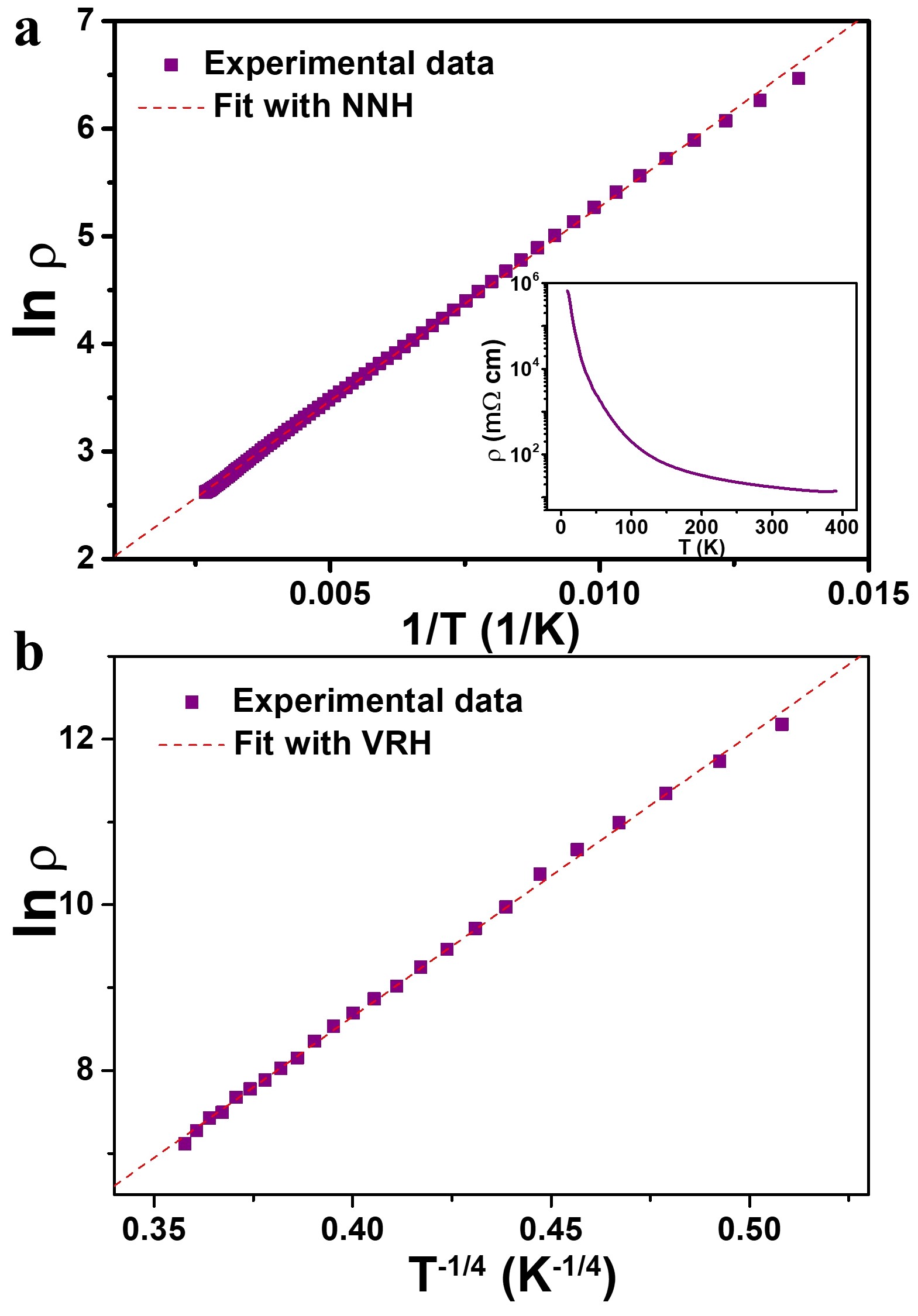}
\caption{\label{fig:onDSO} (a) \textit {ln}$\rho$ \textit{versus} \textit{T}$^{-1}$ for a 10 nm NNO/DSO film, together with the fit to a Near Neighbours Hopping (NNH) model (thermally-activated behaviour) with \textit{E}$_a$=32 meV. A good agreement is found in a large temperature range down to about 70-80K. Inset: Temperature dependence of resistivity in an extended temperature range. (b)  \textit{ln}$\rho$ \textit{versus} \textit{T}$^{-1/4}$ for the temperature range below 70 K, showing the fit to a 3D Variable Range Hopping (VRH) model.} 
\end{figure*}

\renewcommand{\thefigure}{S6} 
\begin{figure*}[h!]
\centering
\includegraphics[scale=0.2]{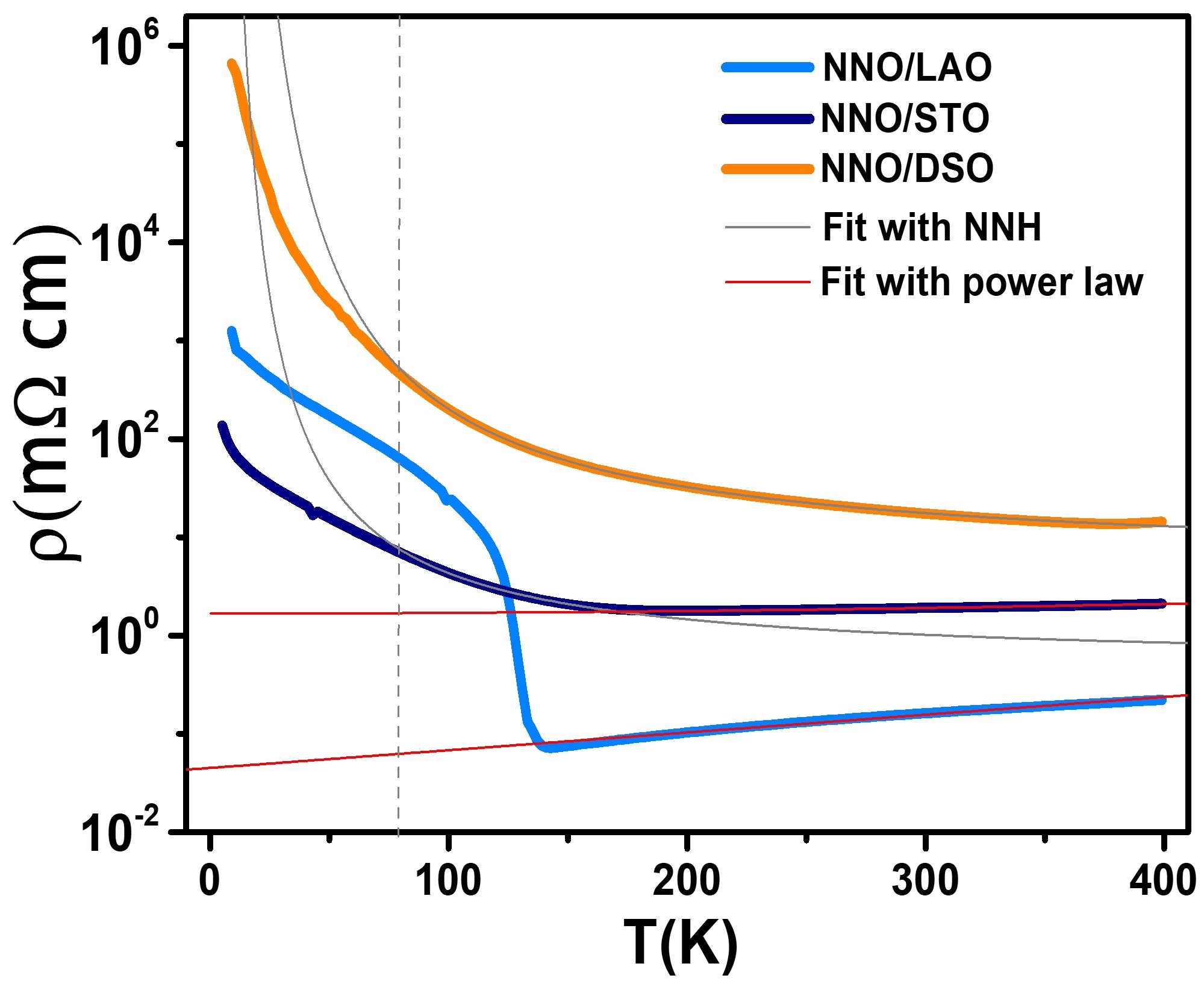}
\caption{\label{fig:s6} $\rho$(T) curves of 10 nm NNO films grown on LAO, STO, and DSO substrates. The vertical dashed line indicates the temperature of crossover from the VRH (fit not shown) to the NNH regimes. The solid grey lines are the best fits in the NNH regime with activation energies of E\textsubscript{a}= 32 meV and E\textsubscript{a}= 20 meV, for the samples on DSO and on STO, respectively. The red solid lines are the best fits with power law in the metallic regime of the NNO films grown on LAO and STO substrates.}
\end{figure*}

\end{document}